\documentclass[prb,aps,twocolumn,showpacs,amsmath,amssymb,10pt]{revtex4-1}
\usepackage{graphicx}
\usepackage{dcolumn}
\usepackage{bm}
\usepackage{amsmath}
\usepackage{epsf}
\usepackage{epstopdf}
\usepackage{epsfig}
\usepackage{graphicx}
\usepackage{braket}
\usepackage{amsfonts}
\usepackage{amsmath}
\usepackage{amssymb}
\usepackage{color,soul}
\usepackage{wasysym}
\usepackage{mathrsfs}
\usepackage{natbib}
\usepackage{multirow}
\usepackage[colorlinks=true,%
linkcolor=blue,%
urlcolor=blue,%
citecolor=blue,%
filecolor=blue,%
bookmarksopen=true,%
pdftitle={SSinQDs},%
pdfsubject={SSinQDs_Manuscript},%
pdfpagemode=UseOutlines]{hyperref}
\DeclareGraphicsExtensions{.pdf,.eps,.png,.jpg,.mps}

\begin{document}

\preprint{APS/123-QED}

\title{Enhancement of the thermoelectric efficiency in a T-shaped quantum dot system in the linear and nonlinear regimes}

\author{G. G\' omez-Silva}
\affiliation{%
Departamento de F\'isica, Pontif\'icia Universidade Cat\'olica do Rio de Janeiro, Caixa Postal 38071,
Rio de Janeiro 22452-970, RJ, Brazil\\
}
\author{P. A. Orellana}
\email{pedro.orellana@usm.cl}
\affiliation{%
Departamento de F\'isica, Universidad T\'ecnica Federico Santa Mar\'ia, Casilla 110V, Valpara\'iso, Chile \\
}
\author{E. V. Anda}
\affiliation{
Departamento de F\'isica, Pontif\'icia Universidade Cat\'olica do Rio de Janeiro, Caixa Postal 38071,
Rio de Janeiro 22452-970, RJ, Brazil\\
}

\date{\today}

\begin{abstract}
In the present work, we investigate the thermoelectric properties of a T-shaped double quantum dot system coupled to two metallic leads incorporating the intra-dot Coulomb interaction. We explore the role of the interference effects and Coulomb blockade on the thermoelectric efficiency of the system in the linear and nonlinear regimes. We studied as well the effect of a Van-Hove singularity of the leads density of states (DOS) at the neighborhood of the Fermi energy, a situation that can be obtained using a carbon nanotube, a graphene nano-ribbon or other contacts with one-dimensional properties. The system is studied above the Kondo temperature. The Coulomb blockade of the electronic charges is studied using the Hubbard III approximation, which properly describes the transport properties of this regime. In the linear response, our results show an enhancement of the thermopower and the figure of merit of the system. For a nonlinear situation, we calculate the thermoelectric efficiency and power output, concluding that the T-shaped double quantum dot is an efficient thermoelectric device. Moreover, we demonstrate the great importance of the DOS Van-Hove singularity at the neighborhood of the Fermi energy to obtain a very significant increase of the thermoelectric efficiency of the system.
\end{abstract}
\maketitle
\section{\label{sec:level1} Introduction}
Thermoelectric effects in low dimensional systems have attracted significant attention in the last decade. Materials with excellent thermoelectric properties can convert heat into electricity (Seebeck effect) or electricity into a temperature gradient (Peltier effect). The performance of a thermoelectric device, in the linear regime, is estimated by the figure of merit $ZT=\mathcal{G}S^2T/\kappa$, where $\mathcal{G}$ is the electronic conductance, $S$ is the thermopower or Seebeck coefficient, $T$ is the temperature and $\kappa$ is the thermal conductivity, which includes contributions from electrons as well as phonons. For practical applications $ZT$ must be as large as possible, so we look for materials characterized by an excellent electronic conductance and at the same time a small thermal conductivity. In bulk materials, these properties are constrained by the well known Wiedemann-Franz (WF) law $L=\kappa/\mathcal{G}T=L_0$, where $L_0=\pi^2k_B^2/3e^2$ is the Lorenz number, $k_B$ the Boltzmann constant and $e$ the electronic charge. This relationship expresses the fact that charge and heat transport are supported by the same scattering processes with a weak dependence of the energy as a consequence of the Fermi liquid theory. Best bulk thermoelectric materials show $ZT<1$, although, to be competitive compared with conventional generators and refrigerators it should be $ZT>3$\cite{Majumdar}. However, nanostructured systems exhibit higher efficiencies than bulk materials according to theoretical predictions \cite{Hicks,Hicks2,Dubi} as well as experiments\cite{Venka,Harman}, which also imply the violation of the WF law\cite{Kubala,Dutta}. One of the phenomena that explains the improvement of the efficiency is the decrease of the thermal conductivity by the increase of the phonon scattering for low dimensional systems\cite{Kithun}. Moreover, Mahan and Sofo\cite{Mahan} show that the efficiency can be improved increasing the density of states (DOS) at the Fermi level. They suggest a maximization of $ZT$ in materials with $\delta$-function form in their DOS. For this reason, quantum dots (QDs) systems are ideal candidates for having a good thermoelectric performance.

The figure of merit $ZT$ is a magnitude calculated in the linear response regime\cite{Hershfield}. This can be understood for bulk materials where the temperature gradient is small inside them, even when the gradient is large throughout the sample. Nevertheless, in nanostructures, especially quantum dots systems, extremely large bias voltage and temperature gradient can be applied. To consider these systems as power generators or cooling devices it is necessary to study the thermoelectric properties in the nonlinear response regime. Due to electronic confinement, transport in nanoscale systems is governed by electron-electron interactions and coherence, which gives rise to interference electronic effects. These are some fundamental ingredients to understand the thermoelectric properties, as it is the case of Fano resonances and Coulomb blockade phenomena.

Thermoelectric properties in systems fabricated with one, two or more QDs have been extensively studied\cite{Boese,Zianni,Swirkowicz,Wierzbicki,Fu,Kennes,Yan,Thierschmann} in different regimes, mostly in the linear regime and, to a lesser extent, nonlinear one. In the linear regime, it was found that interference effects can significantly improve the figure of merit. On the other hand, in the nonlinear regime, several authors \cite{Svensson,Sierra} reported a negative differential thermoconductance, which generates zero thermocurrent at a given temperature gradient. In particular, in the linear regime a T-shaped double quantum dot system was studied in the presence of electron-electron interaction \cite{Hershfield,Monteros,Wojcik,Xu,Wojcik2} using methods as unrestricted Hartree-Fock and Hubbard I (H$_{\text{I}}$) approaches. The system has two possible conduction channels, which allows the observation of interference effects in the conduction. It is important to mention that the same characteristics that increase the figure of merit, could improve the efficiency at a finite bias voltage.

The present paper focuses on the study of the thermoelectric transport through a T-shaped double-quantum-dot (DQD) system coupled to two metallic leads (see Fig. \ref{system}). For a more realistic vision of the problem, we consider an intra-dot Coulomb interaction $U$ in the quantum dots. We explore both, linear and nonlinear regimes. To incorporate the electronic correlations the Hubbard III (H$_{\text{III}}$) approximation is used\cite{Anda}, which provides a reliable description of the Coulomb blockade regime, the transport and thermoelectric properties of the system (see Appendix \ref{app:hubb}). This approximation is a correct approach when the system is above the Kondo temperature. We study as well devices with contacts possessing, in their DOS, Van-Hove singularities at the neighborhood of the Fermi level as it is the case of several one-dimensional systems\cite{Charlier,Hu,Nakada}. We prove this to be an essential ingredient, which significantly enhances the thermoelectric efficiency of these systems.

The paper is organized as follows. In Sec. \ref{sec:model} we describe the model adopted to study the nanosystem. We also outline the H$_{\text{III}}$ approach, as well as the theoretical framework based on the non-equilibrium Green function (NEGF) techniques. In Sec. \ref{sec:res}, we discuss the numerical results obtained, and finally, a summary is given. In Appendix \ref{app:hubb} we conceptually discuss the shortcomings of the H$_{\text{I}}$ approximation.

\section{Model}\label{sec:model}
\begin{figure}
\begin{center}
\centerline{\includegraphics[width=58mm,clip]{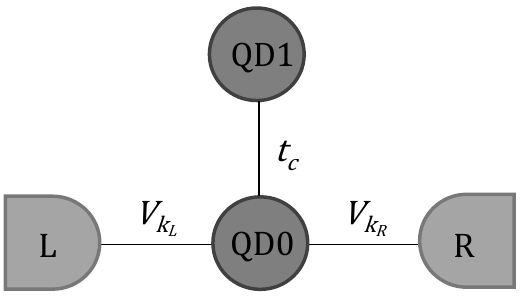}}
\caption{Schematic view of T-shaped DQD system coupled to left ($L$) and right ($R$) metallic leads with an inter-dot coupling denoted by $t_c$.}\label{system}
\end{center}
\end{figure}

We consider a two single level QDs connected to metallic leads, described in Figure \ref{system}. We model the system by a two impurities Anderson Hamiltonian, which can be written as,
\begin{equation}\label{ham:tot}
H=H_{DQD}+H_{leads}+H_{tunnel}.
\end{equation}

\noindent The first term, $H_{DQD}$, describes the DQD molecule, which is given by,
\begin{eqnarray}\label{ham:mol}
H_{DQD}&=&\sum_{i=0,1;\sigma}\varepsilon_{i\sigma}d_{i\sigma}^{\dag}d_{i\sigma}+\sum_{i=0,1}U_in_{i\uparrow}n_{i\downarrow}\nonumber\\
&&+\sum_{\sigma}t_c\left(d_{0\sigma}^{\dag}d_{1\sigma}+\text{H.c.}\right) ,
\end{eqnarray}

\noindent where $\varepsilon_{i\sigma}$ is the level energy of the QD$i$ ($i=0,1$), $d_{i\sigma}$ ($d^{\dag}_{i\sigma}$) is the annihilation (creation) operator of an electron in the QD$i$ with spin index $\sigma$ ($\sigma=\downarrow,\uparrow$), $U_i$ is the local electron-electron interaction energy at QD$i$, $n_{i\sigma}=d_{i\sigma}^\dag d_{i\sigma}$ is the number operator and $t_c$ is the inter-dot tunneling coupling. The second term in Eq. \eqref{ham:tot} describes the electrons in the metallic leads and it is given by,
\begin{equation}\label{ham:lead}
H_{leads}=\sum_{k_\alpha,\sigma}\left(\varepsilon_{k_\alpha\sigma}c_{k_\alpha\sigma}^{\dag}c_{k_\alpha\sigma}+\text{H.c.}\right),
\end{equation}

\noindent where $\varepsilon_{k_\alpha\sigma}$ is the energy of the electron described by the state of quantum number $k_\alpha$ and spin index $\sigma$ at the contact $\alpha$ ($\alpha=L,R$) and $c_{k_\alpha\sigma}$ ($c^{\dag}_{k_\alpha\sigma}$) is the operator that annihilates (creates) it. Finally, the third term is the tunneling Hamiltonian between the leads and the QD0 and is written as,
\begin{equation}\label{ham:tun}
H_{tunnel}=\sum_{k_\alpha,\sigma}\left(V_{k_\alpha}d_{0\sigma}^\dag c_{k_\alpha\sigma}+\text{H.c.}\right),
\end{equation}

\noindent where $V_{k_L}$ ($V_{k_R}$) is the coupling between the embedded QD and the left (right) lead.

The model we propose does not include the inter-dot Coulomb repulsion, which for the parameters taken is at least an order of magnitude less than the intra-dot repulsion. The treatment neglects the splitting between the singlet and triplet configurations of the dots as it is of the order of $t_c^2/U$, an energy value much less than the intra-dot Coulomb repulsion, the dominant many-body interaction in the parameter region where we are studying the system. Besides, no magnetic field is applied so time reversal symmetry is preserved.

To study the physical and in particular the thermoelectric transport properties of this system, we use the NEGF formalism. The intra-dot Coulomb repulsion is treated within the H$_{\text{III}}$ approximation extended to the case of two impurities \cite{Anda}. This approximation correctly describes the electronic and thermoelectric transport in the Coulomb blockade regime. It is important to mention that this is not the case of the H$_{\text{I}}$ approximation\cite{Hubbard} which loses, particularly at resonance, the Coulomb blockade effect in analyzing the transport properties of a system (see Appendix \ref{app:hubb}). The retarded Green function at the QD$0$ is given by,

\begin{equation}\label{green:alph}
G_{00}^r=\sum_{i,j=1}^{2}\frac{p_{0,i}p_{1,j}(\varepsilon-\varepsilon_{1,j})}{(\varepsilon-\varepsilon_{0,i})(\varepsilon-\varepsilon_{1, j})-t_c^2+\text{i}\Gamma(\varepsilon)(\varepsilon-\varepsilon_{1,j})/2},
\end{equation}

\noindent where $p_{0,1}=1-\langle n_0\rangle$, $p_{0,2}=\langle n_0\rangle$, $p_{1,1}=1-\langle n_1\rangle$ and $p_{1,2}=\langle n_1\rangle$. Within the H$_{\text{III}}$ approximation, these quantities can be thought to be the probabilities for the QD0 or QD1 to be single or double occupied with single electronic energies $\varepsilon_{0,1}=\varepsilon_0$, $\varepsilon_{0,2}=\varepsilon_0+U_0$, $\varepsilon_{1,1}=\varepsilon_1$ and $\varepsilon_{1,2}=\varepsilon_1+U_1$, respectively. $\Gamma(\varepsilon)=\Gamma_L(\varepsilon)+\Gamma_R(\varepsilon)$ is the broadening of the dots energy levels due to the connection with the continuum, which is given by $\Gamma_{L(R)}(\varepsilon)=\pi\sum_k\delta(\varepsilon-\varepsilon_{k_{L(R)}})V_{k_{L(R)}}^2$. The density of states (DOS) of the QD is given by $\rho=-(1/\pi)\text{Im}G_{00}^r$. As we can observe from the Eq. \eqref{green:alph}, DOS of the embedded QD has eight poles, two of them for each of the four fractions.

Similarly, we calculate the retarded Green's function at QD1, which can be written as a function of the same probabilities and energies as follows,
\begin{equation}\label{green:beta}
G_{11}^r=\sum_{i,j=1}^{2}\frac{p_{0,i}p_{1,j}[(\varepsilon-\varepsilon_{0,i})+\text{i}\Gamma(\varepsilon)/2]}{(\varepsilon-\varepsilon_{0, i})(\varepsilon-\varepsilon_{1,j})-t_c^2+\text{i}\Gamma(\varepsilon)(\varepsilon-\varepsilon_{1,j})/2}.
\end{equation}

\noindent These Green's functions require a self-consistent calculation to obtain the occupation numbers given by,
\begin{equation}
\langle n_{i}\rangle=\frac{1}{2\pi}\int_{-\infty}^{\infty}G^{<}_{ii}\text{d}\varepsilon,
\end{equation}

\noindent where $G^{<}_{00}=[\Gamma_Lf_L+\Gamma_Rf_R]|G_{00}^r|^2$ and $G^{<}_{11}=[\Gamma_Lf_L+\Gamma_Rf_R]|G_{01}^r|^2$ are the lesser Green's function calculated using the Keldysh formalism, subindex $i$ corresponds to QD sites 0 or 1, $f_{L(R)}=1/[1+\exp{(\varepsilon-\mu_{L(R)})/(k_BT_{L(R)})}]$ is the Fermi-Dirac distribution being $k_B$ the Boltzmann constant, $T_{L(R)}$ the temperature and $\mu_{L(R)}$ the electro-chemical potential corresponding to lead $L(R)$. Finally, $G_{01}^r$ can be expressed as,
\begin{equation}
G_{01}^r=\sum_{i,j=1}^{2}\frac{p_{0,i}p_{1,j}t_c}{(\varepsilon-\varepsilon_{0,i})(\varepsilon-\varepsilon_{1, j})-t_c^2+\text{i}\Gamma(\varepsilon)(\varepsilon-\varepsilon_{1,j})/2}.
\end{equation}

\subsection{Linear response regime}

In the linear response regime, when the temperature gradient and the bias voltage tend to zero, the electric and the heat current, $I$ and $J$, respectively, are given by,
\begin{eqnarray}
I&=&-e^2\mathcal{L}_0V+\frac{e}{T}\mathcal{L}_1\Delta T,\nonumber\\
J&=&e\mathcal{L}_1V-\frac{1}{T}\mathcal{L}_2\Delta T,\\
&&\nonumber
\end{eqnarray}

\noindent where $e$ is the electron charge, $\Delta T$ and $V$ are, respectively, the infinitesimal temperature gradient and applied potential between the contacts and $\mathcal{L}_n$ are the kinetic transport coefficients. They can be calculated integrating the transmission of the system $\tau(\varepsilon)$ as follows\cite{Sivan},
\begin{equation}
\mathcal{L}_n=\frac{2}{h}\int_{-\infty}^{\infty}\left(\frac{\partial f}{\partial\varepsilon}\right)_{\varepsilon=\mu}(\varepsilon-\mu)^n\tau(\varepsilon)\text{d}\varepsilon,
\end{equation}

\noindent where, $h$ is the Planck constant. We obtain the transmission function using the Fisher-Lee relation \cite{FisherLee} $\tau(\varepsilon)=\text{Tr}[\Gamma_LG_{00}^r\Gamma_RG_{00}^a]$ that can be expressed as,
\begin{equation}\label{trans:fun}
\tau(\varepsilon)=-\Gamma(\varepsilon)\text{Im}G_{00}^r.
\end{equation}

\noindent The observable variables can be expressed as functions of the kinetic coefficients. The electronic conductance, at zero temperature gradient, is obtained by $\mathcal{G}=e^2\mathcal{L}_0$. The thermopower, $S=(-1/eT)(\mathcal{L}_1/\mathcal{L}_0)$, is defined as the voltage drop induced by a gradient of temperature when the electric current is zero. The electronic thermal conductance, $\kappa_e=1/T(\mathcal{L}_2-\mathcal{L}_1^2/\mathcal{L}_0)$, is the ratio between the heat current and the temperature gradient when the electric current vanishes. Finally, the thermoelectric efficiency at equilibrium can be described by the figure of merit $ZT=\mathcal{G}S^2T/(\kappa_e+\kappa_{ph})$. The phononic thermal conductance, $\kappa_{ph}$, is neglected in this model. At low temperatures, the thermopower can be obtained using the Mott formula\cite{Jonson}, which is expressed in terms of the electronic conductance, and it is given by $S=(\pi^2/3)(k_B^2T/e)(\text{d}\ln\mathcal{G}/\text{d}\varepsilon)_{\varepsilon=\mu}$. However, this formula is no longer valid\cite{Gomez-Silva} at
the presence of Fano antiresonances that causes the conductance to vanish, which would imply a divergence of the thermopower.

\subsection{Nonlinear regime}

In the nonlinear regime, the electric current can be written as,
\begin{equation}
I=\frac{e}{h}\int_{-\infty}^{\infty}\tau(\varepsilon)[f_L(\varepsilon)-f_R(\varepsilon)] \text{d}\varepsilon.
\end{equation}

\noindent We also can derive an expression for the heat current using the first law of thermodynamics
\begin{equation}\label{firstlaw}
\text{d}U_\alpha=\text{d}Q_\alpha+\text{d}W_\alpha,
\end{equation}

\noindent where $\alpha=L,R$; $\text{d}W_\alpha=\mu_\alpha\text{d}N_\alpha$ is the work done by the reservoir $\alpha$ and d$Q_\alpha$ is the transmitted heat between the reservoirs. We write the rate of change of the quantities in Eq. \eqref{firstlaw} as $J_E=J_\alpha+\mu_\alpha I$. Finally, the expression for the heat current is\cite{Yamamoto},
\begin{equation}
J_\alpha=\frac{1}{h}\int_{-\infty}^{\infty}(\varepsilon-\mu_\alpha)\tau(\varepsilon)[f_L(\varepsilon)-f_R(\varepsilon)] \text{d}\varepsilon.
\end{equation}

In this regime, we can regard this system as a heat engine. We set $\mu_R$ higher than $\mu_L$. The work done by the reservoir per unit of time is equivalent to the power output $dW/dt=P=IV$, with $V=(\mu_R-\mu_L)/e$. The efficiency is defined as the ratio between the work done and the heat current extracted from the high-temperature reservoir $\eta=P/J_L$, per unit of time. Therefore,

\begin{equation}
\eta=\frac{(\mu_R-\mu_L)\int_{-\infty}^{\infty}\tau(\varepsilon)[f_L(\varepsilon)-f_R(\varepsilon)]\text{d}\varepsilon} {\int_{-\infty}^{\infty}(\varepsilon-\mu_L)\tau(\varepsilon)[f_L(\varepsilon)-f_R(\varepsilon)]\text{d}\varepsilon}.
\end{equation}

\section{Results}\label{sec:res}

\subsection{Linear Response}\label{lin:res}

In this section, we discuss the thermoelectric properties at finite temperature ($T\ne0$). We consider two types of leads. First, we consider normal metallic leads where the wide band limit can be used by taking a constant broadening $\Gamma_{L(R)}$ ($\Gamma_0=\Gamma_L+\Gamma_R$), where we take $\Gamma_0$ as the energy unit. On the other hand, it is known that quasi-1D systems, as it is the case of carbon nanotubes or graphene nanoribbons, exhibit Van-Hove singularities in their DOS. These singularities can be taken to be near the Fermi level. In this second case, we assume that the connection with the continuum $\Gamma_{L(R)}(\varepsilon)=\pi V_{k_{L(R)}}^2\rho(\varepsilon)$ is no longer constant and exhibits a Van-Hove singularity. The 1D leads DOS, $\rho(\varepsilon)$, can be represented around the Fermi level as,
\begin{equation}
\rho(\varepsilon)=\begin{cases}A/\sqrt{\varepsilon-\varepsilon_{VH}},\,\qquad \text{if $\varepsilon>\varepsilon_{VH}$},\\
B,\quad\qquad\qquad\qquad \text{if $\varepsilon\leq\varepsilon_{VH}$},
\end{cases}
\end{equation}
where $A$ and $B$ are constants that depend on the geometry of the leads and $\varepsilon_{VH}$ is the energy where the Van-Hove singularity is localized. For metallic and semiconductors carbon nanotubes, $A$ and $B$ have been explicitly calculated \cite{Mintmire}. For the sake of simplicity, we take $A=1/\pi \sqrt{2D}$ and $B=1/\pi D'$, being $D$ and $D'$ the bandwidth of two different bands, as it would be the case of a nanoribbon. We take the energy unit as the coupling value $\Gamma_0=\Gamma_{L(R)}(\varepsilon_{F})$. Besides, the Fermi level of the 1D contact can be tuned doping the material\cite{Kim,Kongkanand}.

Other parameters to be considered in this section are the inter-dot coupling $t_c=2\Gamma_0$, the temperature of the leads $k_BT_L=k_BT_R=0.1\Gamma_0$ and the bias voltage $\Delta\mu=(\mu_L-\mu_R)\rightarrow0$. Besides, we consider the local Coulomb repulsion satisfying $U_0=U_1=U$, symmetric couplings to the leads, $\Gamma_{R}=\Gamma_{L}=\Gamma_0$ and $\varepsilon_F=0$. We choose a set of parameters ($T$, $t_c$ and $U$) ensures that the system is above the Kondo temperature ($T_K=\sqrt{\Gamma U}\exp[-\pi|\varepsilon_0||\varepsilon_0+U|]$), for a gate potential $-U/2<\varepsilon_0<-2\Gamma_0$, where the system would be at the Kondo regime.

\begin{figure*}
\centerline{\includegraphics[width=88mm,height=73mm,clip]{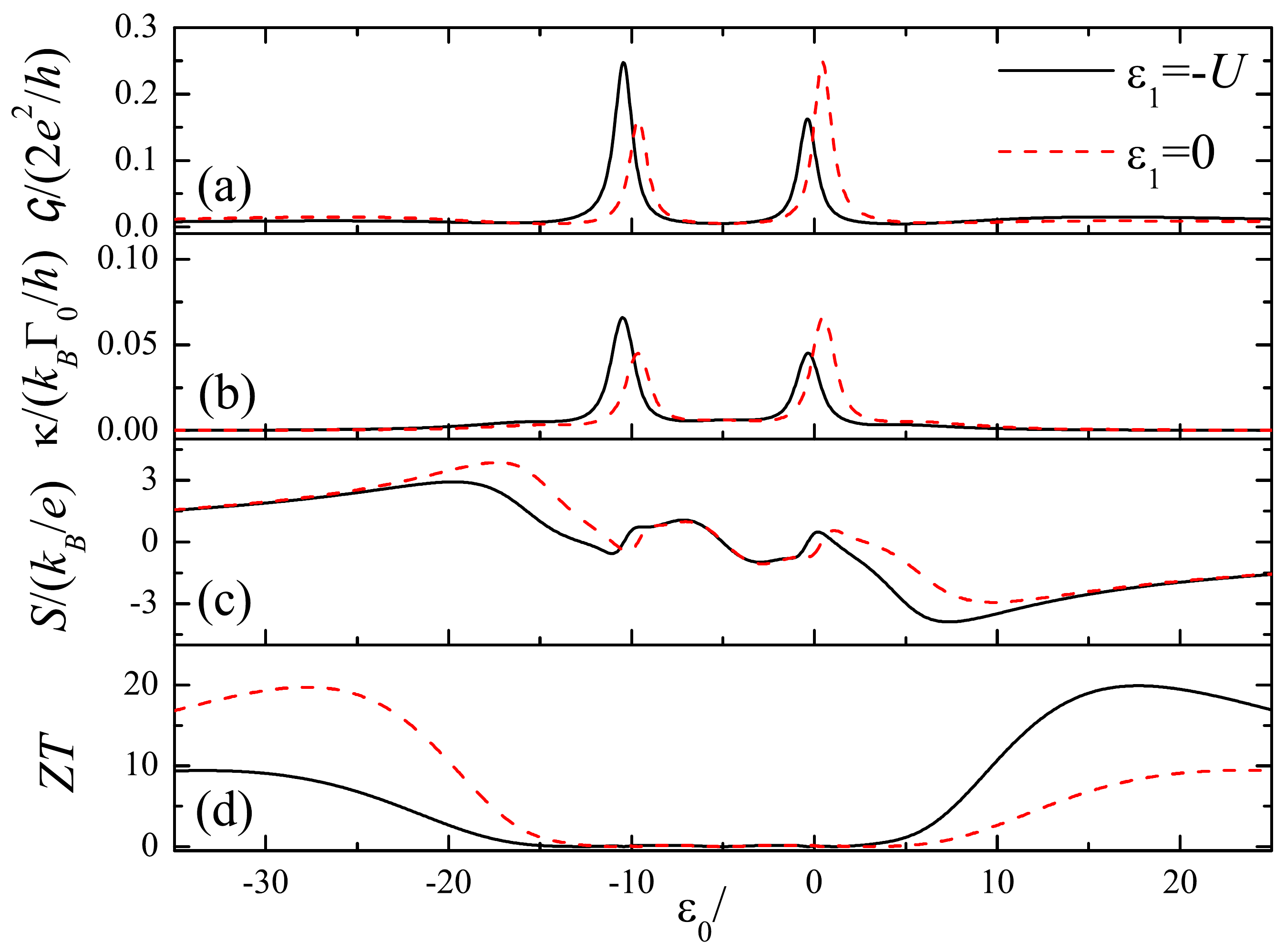}\hspace{0.2cm}\includegraphics[width=88mm,height=73mm,clip]{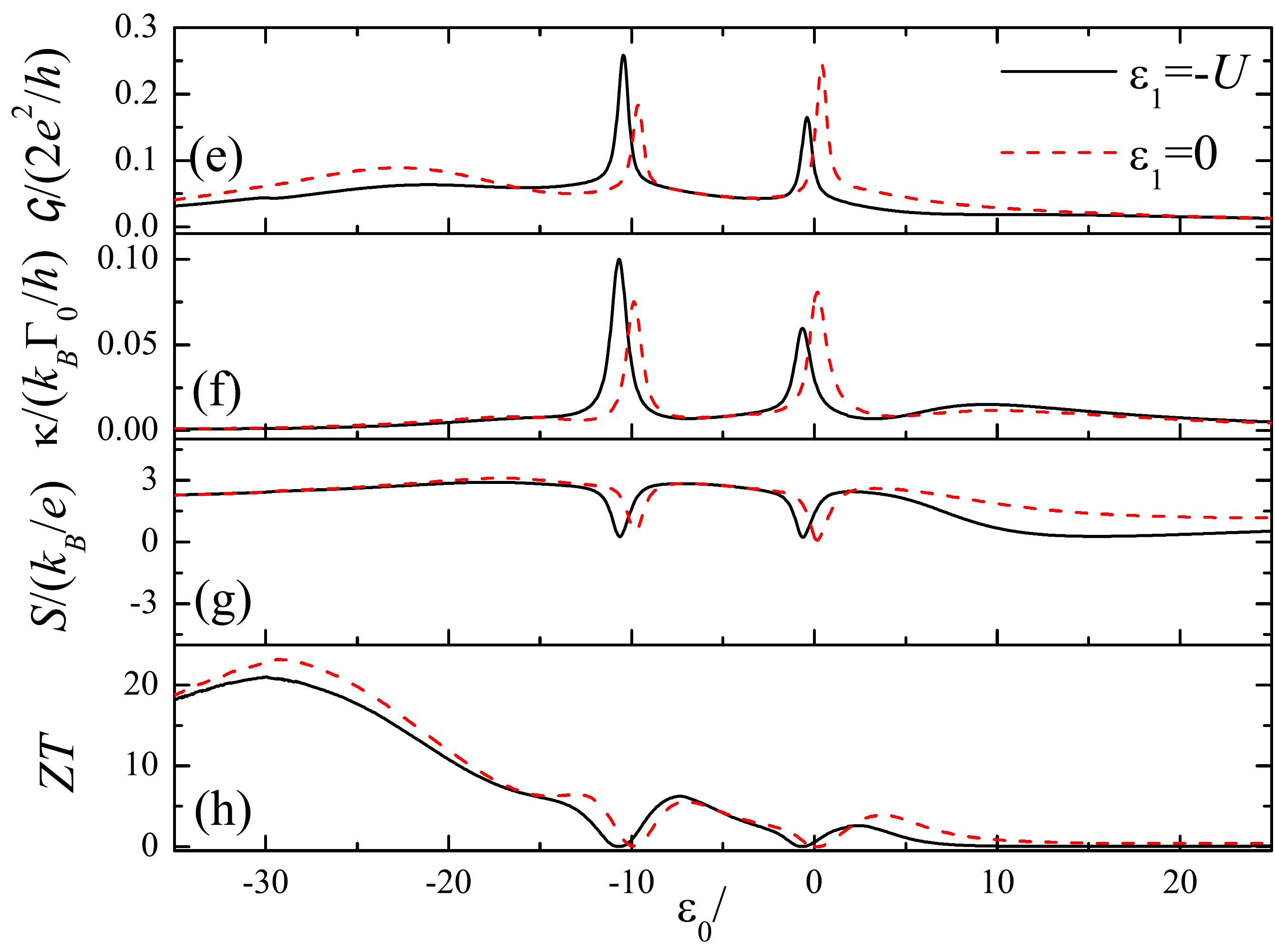}}
\caption{(a) and (e) Electronic conductance, (b) and (f) thermal conductivity, (c) and (g) Seebeck coefficient, and, (d) and (h) figure of merit as a function of the gate potential $\varepsilon_0$ for $t_c=2\Gamma_0$, $k_BT=0.1\Gamma_0$, $U=10\Gamma_0$ and two different values of the gate potential $\varepsilon_1$. Left and right panels correspond to normal and 1D leads, respectively, and in the last case $\varepsilon_{VH}=0.2\Gamma_0$.}\label{equil1}
\end{figure*}

Figure \ref{equil1} displays the electronic conductance $\mathcal{G}$, thermal conductivity $\kappa$, thermopower $S$ and figure of merit $ZT$ as a function of the embedded QD gate voltage $\varepsilon_1$, for normal (left panels) and 1D (right panels) leads. We set the gate voltage of the side coupled QD in two different values that correspond to the resonance energies $\varepsilon_1=-U$ (black solid line) and $\varepsilon_1=0$ (red dashed line). We observe, in Figure \ref{equil1} (a) and (b), two peaks in the linear conductance and thermal conductivity, respectively, when $\varepsilon_0$ is as well at resonance with the Fermi level. We could expect the appearance of eight peaks corresponding to the eight poles of the Eq. \eqref{green:alph}. However, when $\varepsilon_1$ is fixed at resonance, only two fractions in $G_{00}^r$ depend on $\varepsilon_0$ and, therefore, only two peaks are observed in the conductance. The same peaks can be observed in panels (e) and (f), when the system is connected with 1D type of leads. We note that in this case, in the off resonance region, the values of $\mathcal{G}$ and $\kappa$ are larger than for the system connected to normal leads. This behavior can be attributed to the Van-Hove singularity present in the leads DOS. We see, in Figure \ref{equil1} (c) and (g), that the maxima values of thermopower are essentially the same for the two types of leads. However, the thermopower for 1D leads takes higher values between the resonances. As it will be discussed below, this result is relevant for the nonlinear regime. The figure of merit for normal leads assumes small values close to the embedded QD resonances, as shown in Figure \ref{equil1} (d). When we take $\varepsilon_0$ far from the resonances, there is a significant increase of $ZT$. We note that when $\varepsilon_1=-U$ (black solid line), the high enhanced of $ZT$ occurs for positive energies, whereas when $\varepsilon_1=0$, it occurs for negative energies. This opposite situations can be explained as follows. It is known that the figure of merit increases due to abrupt changes in the transmission function, which is proportional to the DOS, as shown in Eq. \eqref{trans:fun}. For this system, there is always a projection of the side coupled QD local levels onto the embedded QD DOS. This projection generates a resonance in the transmission with a broadening which is inversely proportional to the difference between $\varepsilon_0$ and $\varepsilon_1$. So, the larger the difference between the gate potentials, the narrower the broadening of the resonance for the side coupled dot. A narrow resonance implies an abrupt change in the transmission and consequently an enhancement of the figure of merit. When the difference between $\varepsilon_0$ and $\varepsilon_1$ is large enough, this effect on the transmission reduces and the figure of merit begins to decrease. Panel (h) of Figure \ref{equil1} shows the figure of merit for 1D leads. We observe an increase of $ZT$ even for the region close to the embedded QD resonance. In this case, it is the projection of the Van-Hove singularity onto the embedded QD DOS that enhances $ZT$. In addition, for normal leads we observe that the curves for $\varepsilon_1=-U$ and $\varepsilon_1=0$ are symmetric around $\varepsilon_0=-U/2$. For 1D leads, this symmetry is broken.

\begin{figure*}
\centerline{\includegraphics[width=88mm,clip]{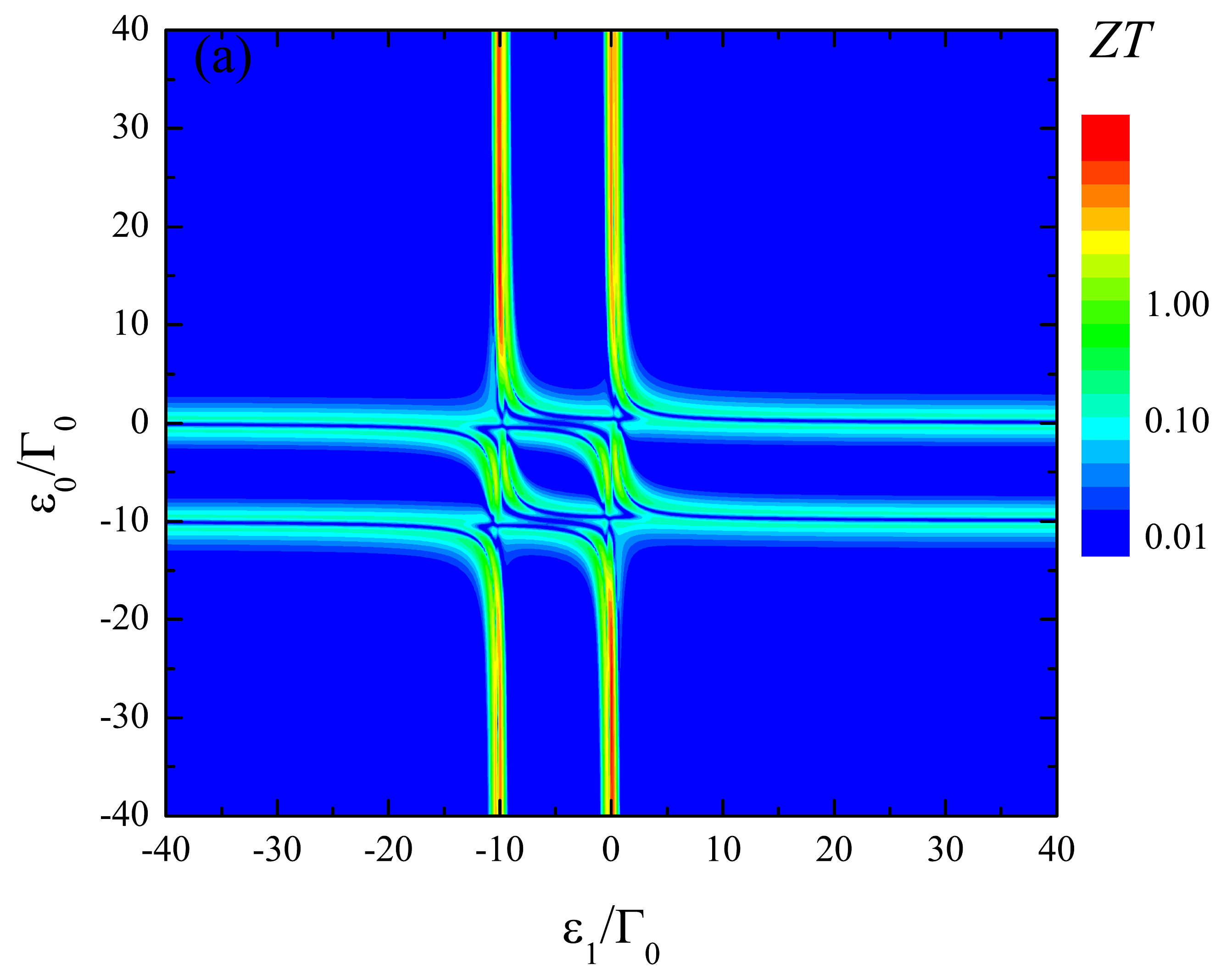}\includegraphics[width=88mm,clip]{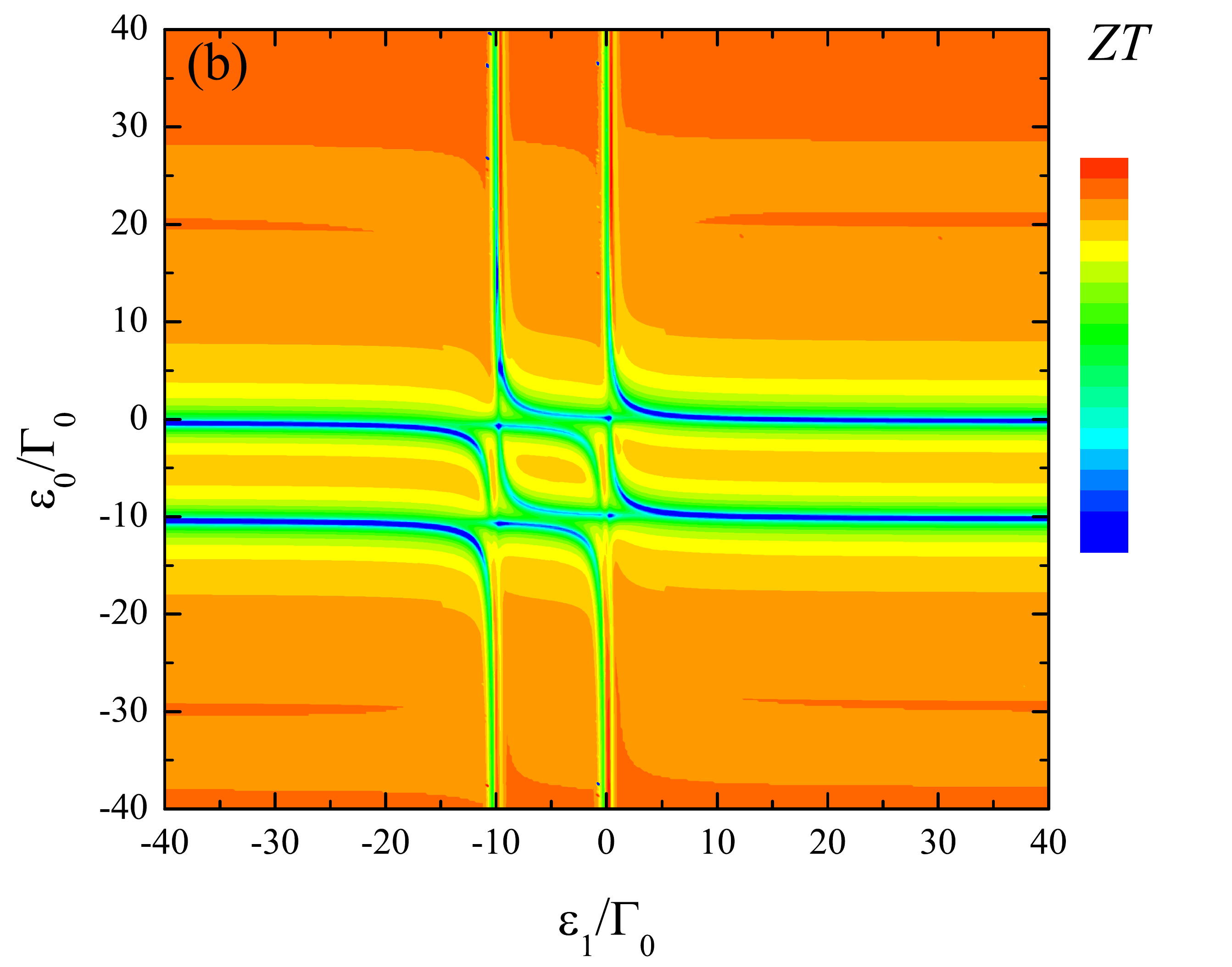}}
\caption{The contour plots display the figure of merit ZT in log scale as a function of gate potentials $\varepsilon_{0}$ and $\varepsilon_{1}$ for (a) normal and (b) 1D leads, $t_c=2\Gamma_0$, $k_BT=0.1\Gamma_0$, $U=10\Gamma_0$ and $\varepsilon_{VH}=0.2\Gamma_0$.}\label{ZT}
\end{figure*}

In the previous analysis we have studied the equilibrium thermoelectric properties of the system assuming the side coupled dot to be at resonance, $\varepsilon_1=-U$ and $\varepsilon_1=0$. We are now studying the figure of merit $ZT$ in the parameter space. Figure \ref{ZT} displays $ZT$ (in a log scale) for all values of gate potentials $\varepsilon_{0}$ and $\varepsilon_{1}$, in the case of normal (left panel) and 1D (right panel) leads. For normal leads (Figure \ref{ZT} (a)), we identify two regions of high values of $ZT$; the central region with both QDs at resonance and another region with the side-coupled QD at resonance and the embedded QD off resonance. In the last case, $ZT$ reaches its maxima values but in a narrow region of $\varepsilon_1$. When both, embedded and side-coupled QDs, are off resonance, $ZT$ is essentially zero and it is possible to observe a splitting of the peaks at the central region. It is important to mention that, even when these values of $ZT$ are low, they are placed in a region where the linear conductance take non-zero values. We will discuss the significance of the relationship between efficiency and conductance in the section \ref{efficiency}. Alternatively, for 1D leads (Figure \ref{ZT}(b)), the enhancement of the figure of merit appears in all the $\varepsilon_{0}$, $\varepsilon_1$ parameter space, except when the embedded QD is at resonance. Here, the contact Van-Hove singularity close to the Fermi level produces a very significant enhancement of the figure of merit. This remarkable result implies that the presence of the Van-Hove singularity by itself is very effective to increase the efficiency. Furthermore, as it was as well the case of normal leads, the narrow region where the embedded QD is off resonance and the side-coupled QD is at resonance shows the highest values of $ZT$. Even though the Van-Hove singularity by itself increases the efficiency, the effect of connecting a side-coupled QD at resonance enhances this effect.

\begin{figure}
\centerline{\includegraphics[width=88mm,height=73mm,clip]{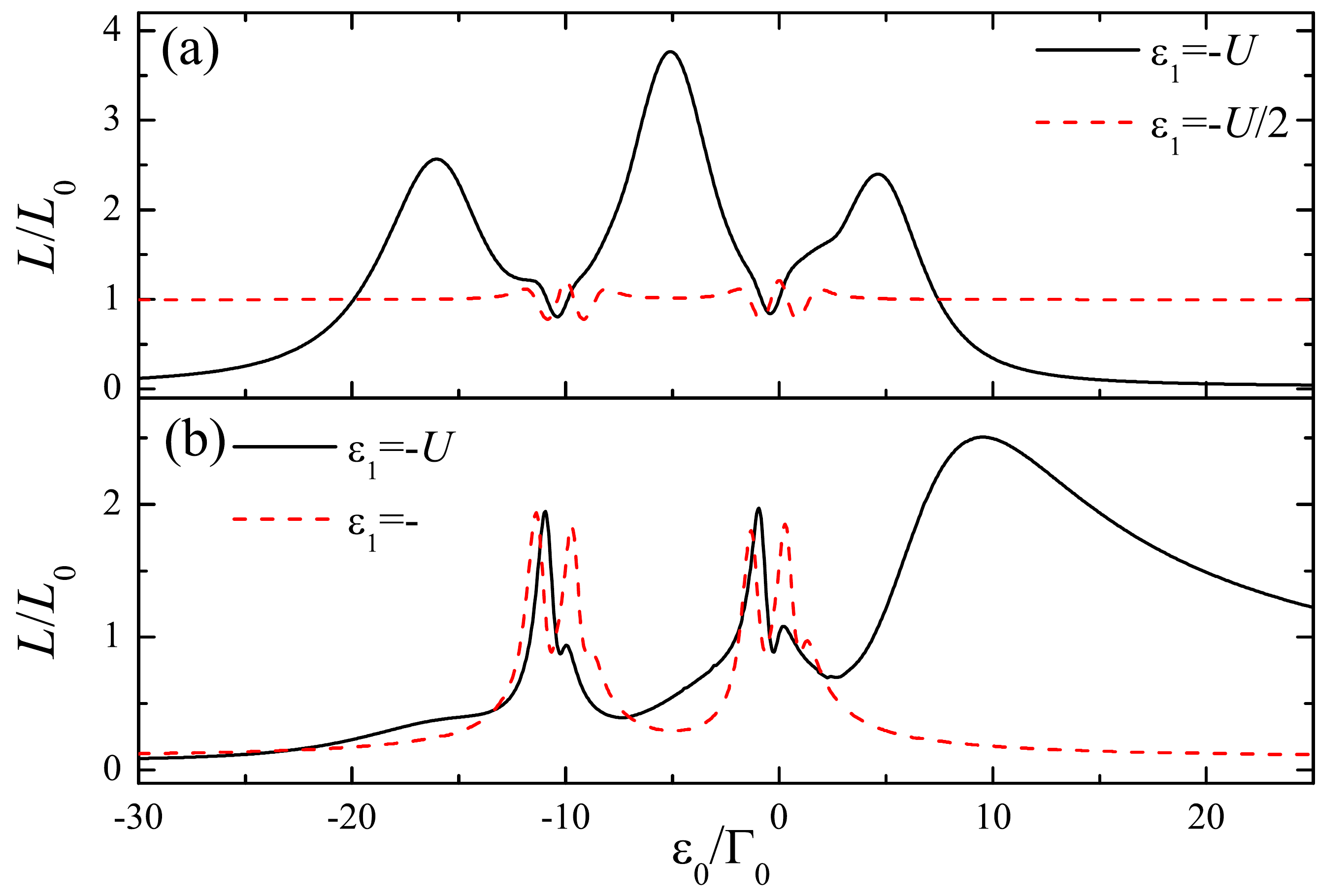}}
\caption{Lorentz number as a function of the gate potential $\varepsilon_0$ for $t_c=2\Gamma_0$, $k_BT=0.1\Gamma_0$, $U=10\Gamma_0$ and two different values of the gate potential $\varepsilon_1$. (a) and (b) panels correspond to normal and 1D leads, respectively.}\label{equil3}
\end{figure}

Figure \ref{equil3} displays the Lorentz number for (a) normal and (b) 1D leads, for two values of the gate potential $\varepsilon_1$. In the panel (a) we can appreciate that for $\varepsilon_1=-U$ (black solid line, the side-coupled dot at resonance) the Wiedemann-Franz law is violated in almost all range of $\varepsilon_0$, although, it is not the case for $\varepsilon_1=-U/2$ (red dashed line, the side-coupled dot off resonance) where the Wiedemann-Franz law holds in almost all values of $\varepsilon_0$. The case of panel (b) is different. Here we observe that the Wiedemann-Franz law is violated for all values of $\varepsilon_0$ for the side-coupled dot being or not at resonance. For normal leads, we see that it is the resonance condition of the QDs that controls the fulfillment or not of the law (observe that it is violated as well when $\varepsilon_0=-U$ and $\varepsilon_0=0$). As it was the case of the enhancement of $ZT$, it is the projection of the states of the side-coupled QD onto the embedded QD that generates a narrow resonance and violates the law. For 1D leads, a narrow resonance is always present around the Fermi level due to the Van-Hove singularity of the leads DOS, and therefore the law is violated independently of the values of $\varepsilon_0$ and $\varepsilon_1$. As it expected, the strong dependence of the electronic conductance and thermal conductivity with the Fermi energy causes a strong violation of the Wiedemann-Franz law.

\subsection{Non-linear regime}\label{nonlin:res}

\begin{figure}
\centerline{\includegraphics[width=88mm,height=73mm,clip]{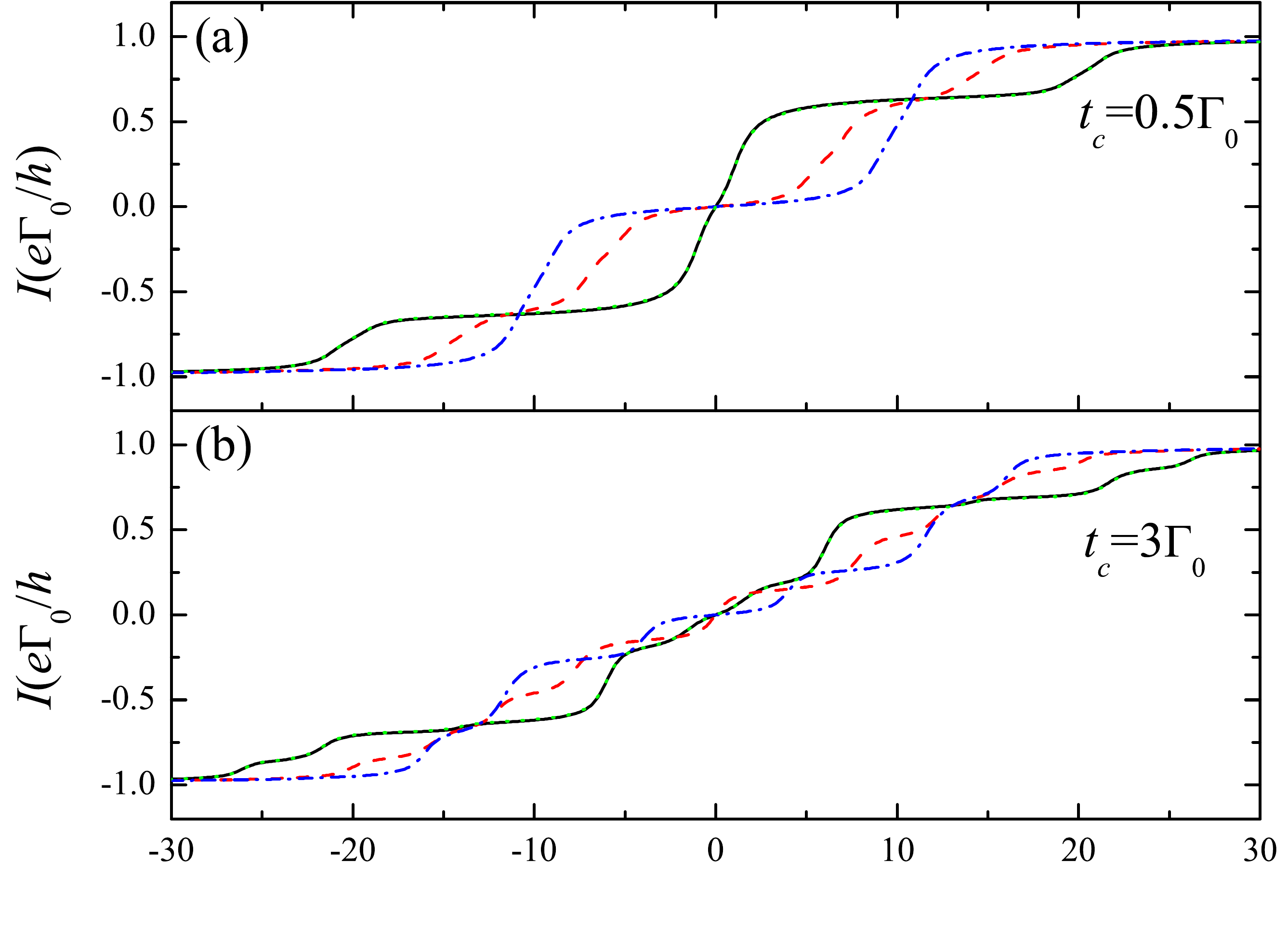}}
\caption{Electrical current as a function of the bias voltage $eV$ for different values of the local energies $\varepsilon_d$ at zero temperature gradient, $\Delta T =0$. Other parameters are $k_BT=0.1\Gamma_0$ and, $t_c=0.5\Gamma_0$ for (a) and $t_c=3\Gamma_0$ for (b) panel, respectively. The values of the local energies are $\varepsilon_d=0$ (black solid line), $\varepsilon_d=-3U/10$ (red dashed line), $\varepsilon_d=-U/2$ (blue dash-dotted line) and $\varepsilon_d=-U$ (green dotted line).}\label{noneq1}
\end{figure}

In this section, we explore the thermoelectric properties in a nonlinear regime present when the system is under the effect of a finite bias voltage $V$ and a temperature gradient $\Delta T$. We consider $T$ as a background temperature being $T_R=T$ and $T_L=T+\Delta T$. Similarly, we set the chemical potential of the leads $\mu_L=eV/2$ and $\mu_R=-eV/2$, $U=10\Gamma_0$ and we take four different values of the local QD energies $\varepsilon_0=\varepsilon_1=\varepsilon_d$. Figure \ref{noneq1} (a) shows the electric current as a function of the bias voltage for a inter-dot coupling $t_c=0.5\Gamma_0$. In this case, the local state of the embedded dot predominates because the inter-dot coupling is weak and we see, for all cases, a characteristic plateau of the Coulomb blockade regime, where there is an increase in the current when the local levels align with the chemical potential of the leads. The same curves are represented in the Figure \ref{noneq1} (b) for $t_c=3\Gamma_0$. Here, we observe several increases in the current generated by the larger coupling to side-coupled QD. The connection with this QD gives to the system more channels for the current to go along. In this case, we observe Ohm's law behavior. The figure shows that the current flowing between the leads is zero when no voltage is applied. In both panels, the curves for $\varepsilon_d=0$ and $\varepsilon_d=-U$ are equivalent because the QDs are at resonance with a Fermi energy, $\varepsilon_F=0$.

\begin{figure}
\centerline{\includegraphics[width=88mm,height=73mm,clip]{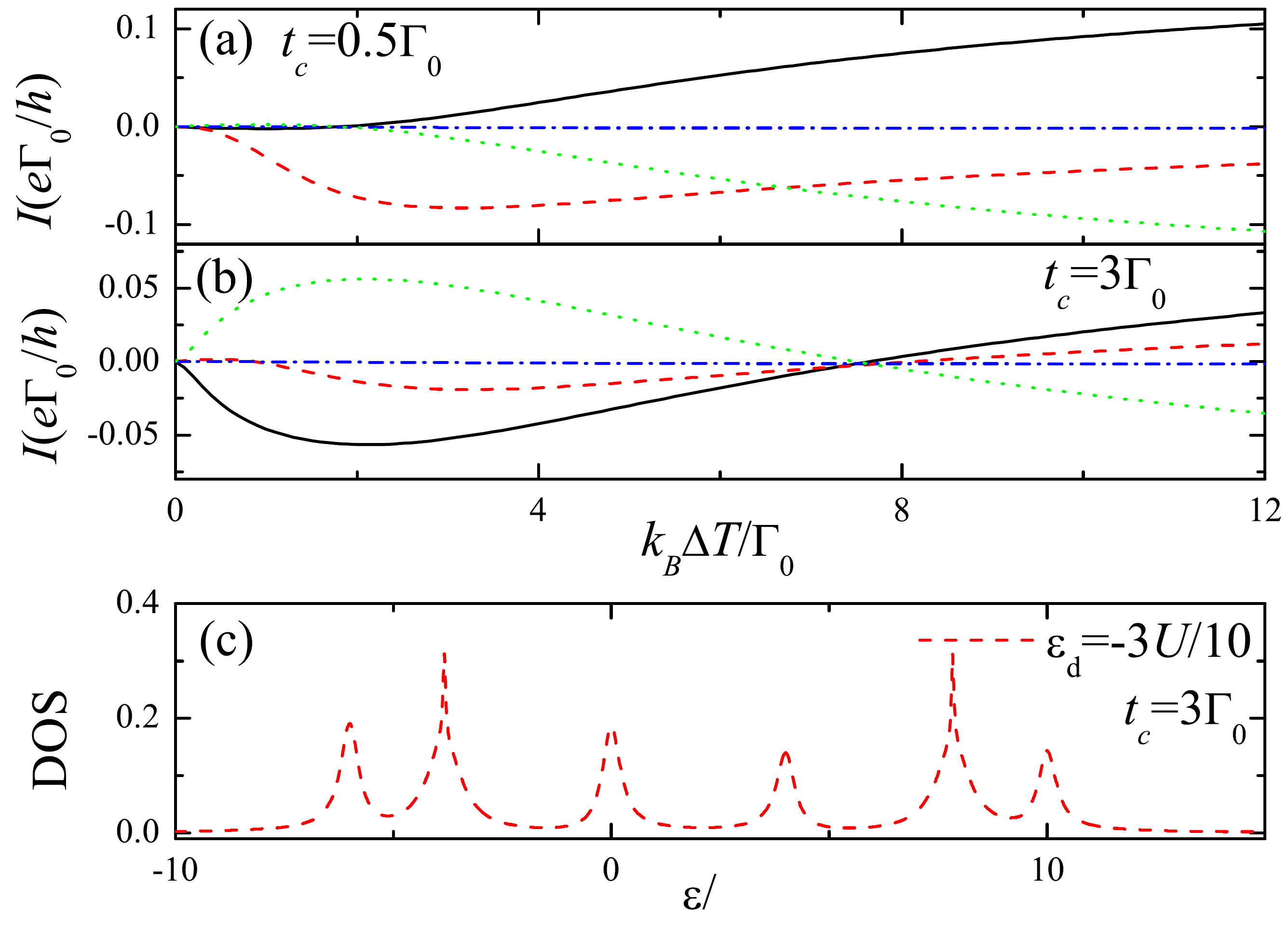}}
\caption{Electrical current as a function of the temperature gradient $k_B\Delta T$ for different values of the local energies $\varepsilon_d$ at zero bias voltage. Other parameters are $k_BT=0.1\Gamma_0$, and, $t_c=0.5\Gamma_0$ for (a) and $t_c=3\Gamma_0$ for (b) panel, respectively. The values of the local energies are $\varepsilon_d=0$ (black solid line), $\varepsilon_d=-3U/10$ (red dashed line), $\varepsilon_d=-U/2$ (blue dash-dotted line) and $\varepsilon_d=-U$ (green dotted line). Panel (c) shows the DOS of the embedded QD as a function of the energy for $k_BT=0.1\Gamma_0$ and $t_c=3\Gamma_0$.}\label{noneq2}
\end{figure}
\begin{figure}
\centerline{\includegraphics[width=88mm,height=73mm,clip]{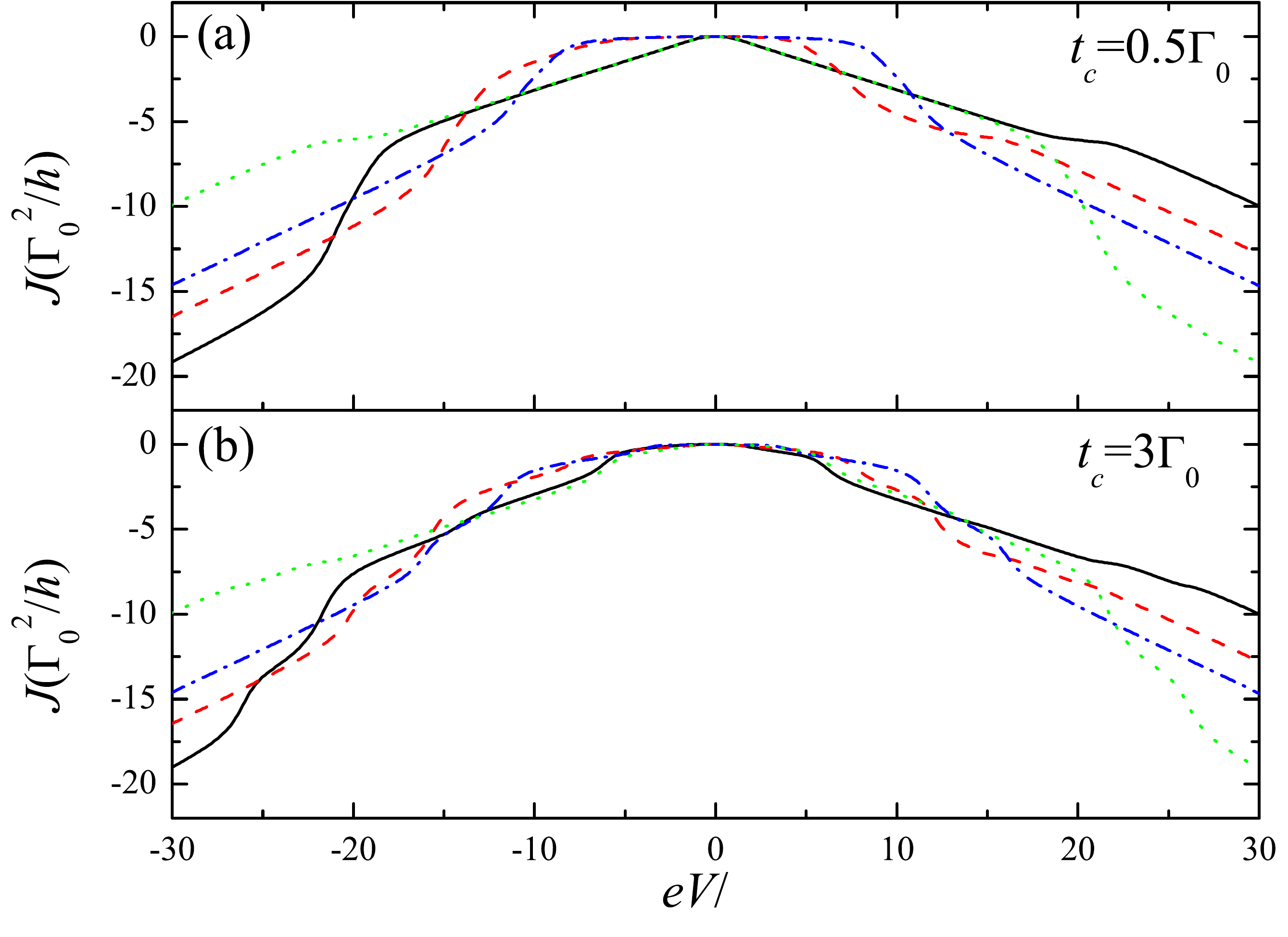}}
\caption{Heat current as a function of the bias voltage $eV$ for different values of the local energies $\varepsilon_d$ at zero temperature gradient. Other parameters are $k_BT=0.1\Gamma_0$, and, $t_c=0.5\Gamma_0$ for (a) and $t_c=3\Gamma_0$ for (b) panel, respectively. The values of the local energies are $\varepsilon_d=0$ (black solid line), $\varepsilon_d=-3U/10$ (red dashed line), $\varepsilon_d=-U/2$ (blue dash-dotted line) and $\varepsilon_d=-U$ (green dotted line).}\label{noneq3}
\end{figure}

In Figure \ref{noneq2}, we explore the current as a function of the temperature gradient at zero bias for the same local energies and inter-dot couplings as before. As expected, for a weak inter-dot coupling, $t_c=0.5\Gamma_0$ (see Figure \ref{noneq2} (a)), we see a nonlinear behavior of the current. For $\varepsilon_d=-U/2$, the current is always zero because the DOS is symmetric around the Fermi energy $\varepsilon_F=0$. For $\varepsilon_d=-U$ we observe a little increase of the thermocurrent with $\Delta T$. However, when we continue heating one lead, it reaches a maximum, decreases and even changes its sign. This situation was already studied and explained by Sierra \emph{et al.}\cite{Sierra}. For $t_c=3\Gamma_0$ (see Figure \ref{noneq2} (b)), the curve for $\varepsilon_d=-3U/10$ (red dashed line) shows that the thermocurrent is positive, then it changes to negative values and finally it becomes positive as we increase $\Delta T$. This behavior can be explained looking at the embedded QD DOS in Figure \ref{noneq2} (c). The temperature gradient is applied taking $\varepsilon_F=0$. Then the states with positive energy contribute to the current with electronic carriers while the negative energy carriers are holes. Since the DOS is not symmetric with respect to $\varepsilon_F$ due to the side-coupled QD and the Coulomb interaction, the current changes its sign twice as we increase $\Delta T$. Similar results were found in a parallel coupled double quantum dot system\cite{Sierra2}.

Figure \ref{noneq3} shows the heat current (considering $J\equiv J_L$) for different values of $\varepsilon_d$ and two different values of $t_c$. We observe that $J$ is symmetric around $V=0$ only for the particle-hole symmetry point $\varepsilon_d=-U/2$. Moreover, we see an invariance between $J(eV)$ for $\varepsilon_d=0$ and $J(-eV)$ for $\varepsilon_d=-U$, i. e., when we invert $\varepsilon_d$ around the particle-hole symmetry point and change $eV$ for $-eV$. The same invariance is observed for a single QD\cite{Sierra3}. It is also clear the nonlinearity of the process for both $t_c=0.5\Gamma_0$ and $t_c=3\Gamma_0$, however, the heat current depends linearly on $eV$ when $\varepsilon_d$ is far for the Fermi energy of the leads, which is easier to observe when $t_c=0.5\Gamma_0$.

\subsection{Efficiency calculation}\label{efficiency}
\begin{figure*}[ht]
\centerline{\includegraphics[width=88mm,height=73mm,clip]{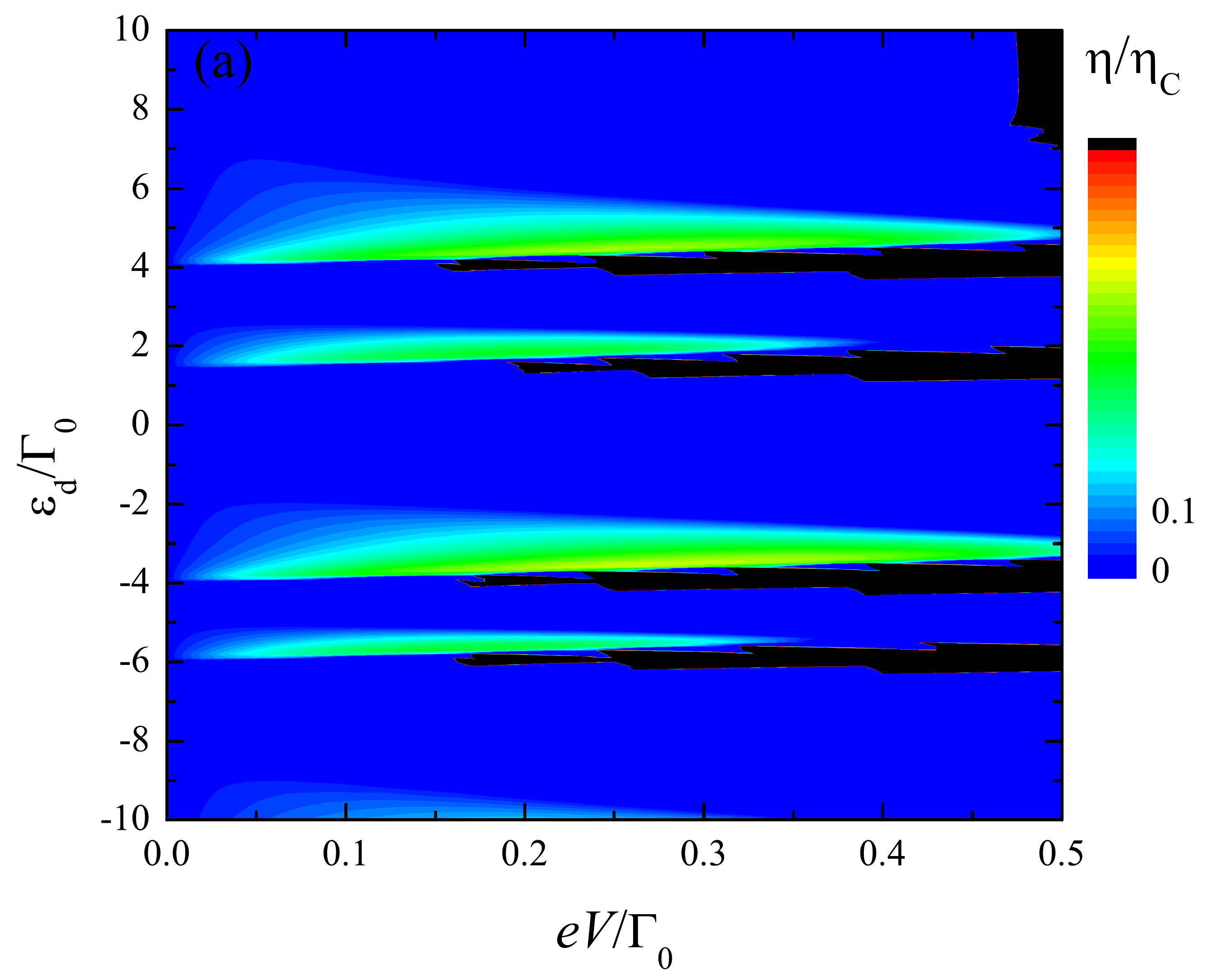}
\includegraphics[width=88mm,height=73mm,clip]{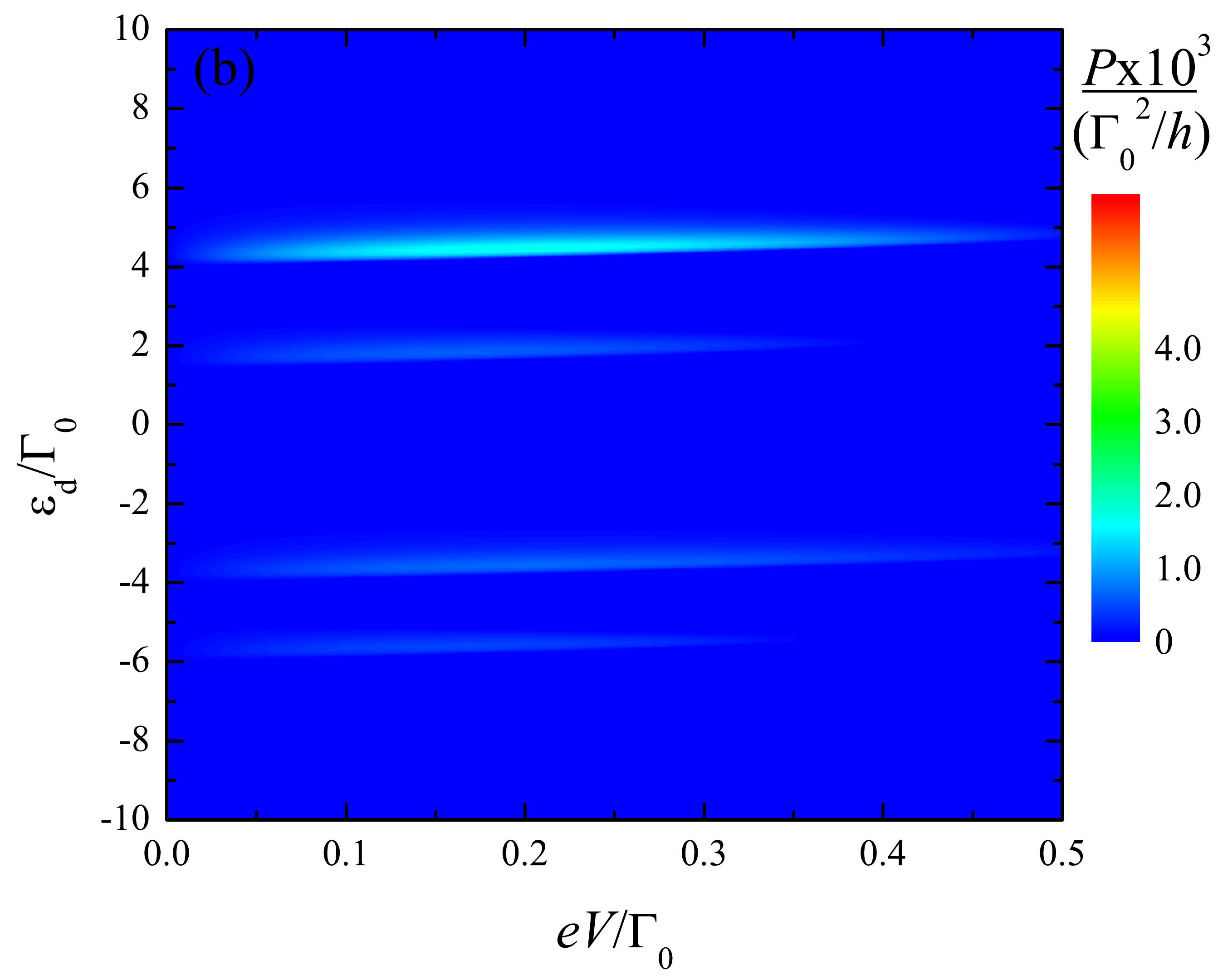}}
\centerline{\includegraphics[width=88mm,height=73mm,clip]{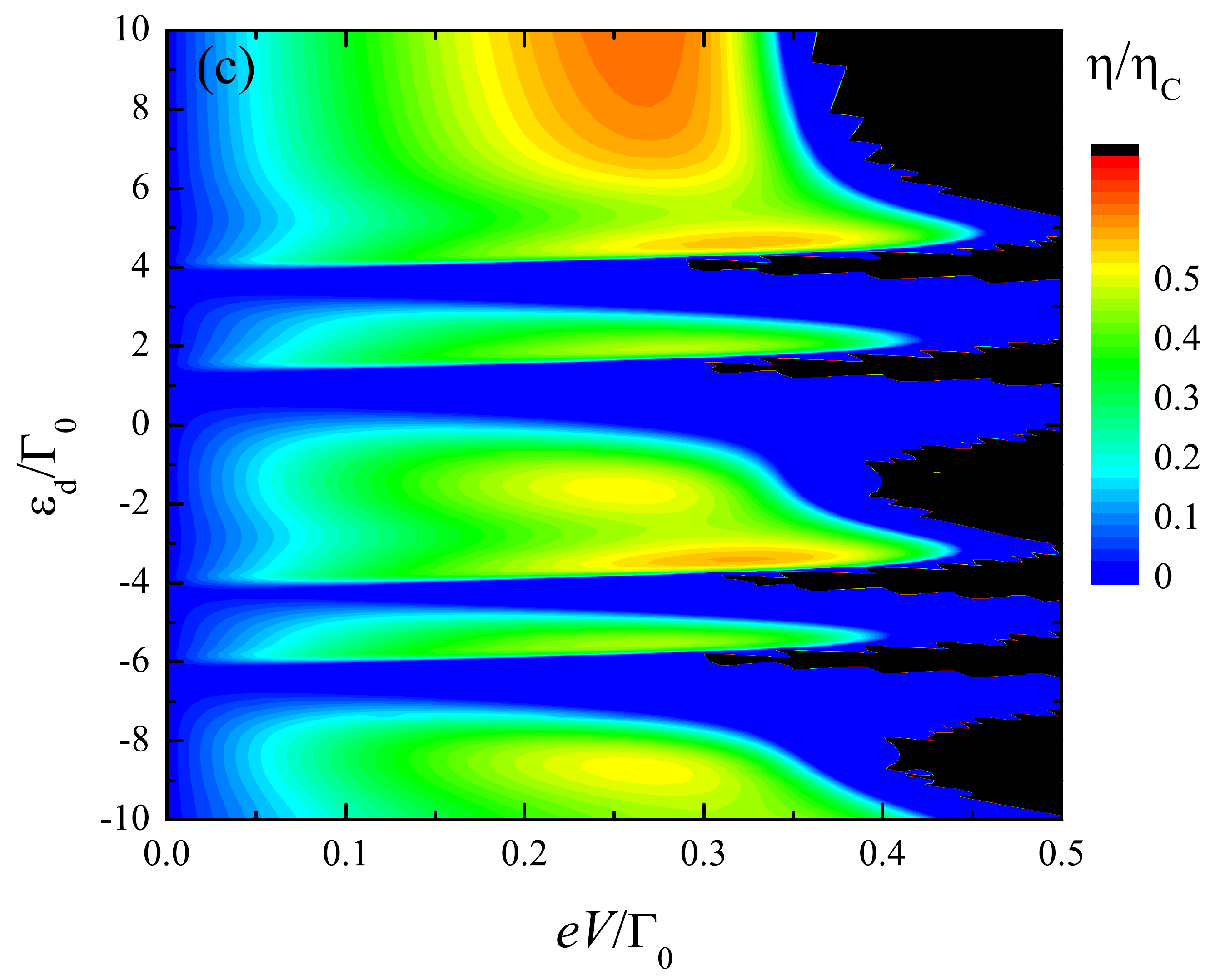}
\includegraphics[width=88mm,height=73mm,clip]{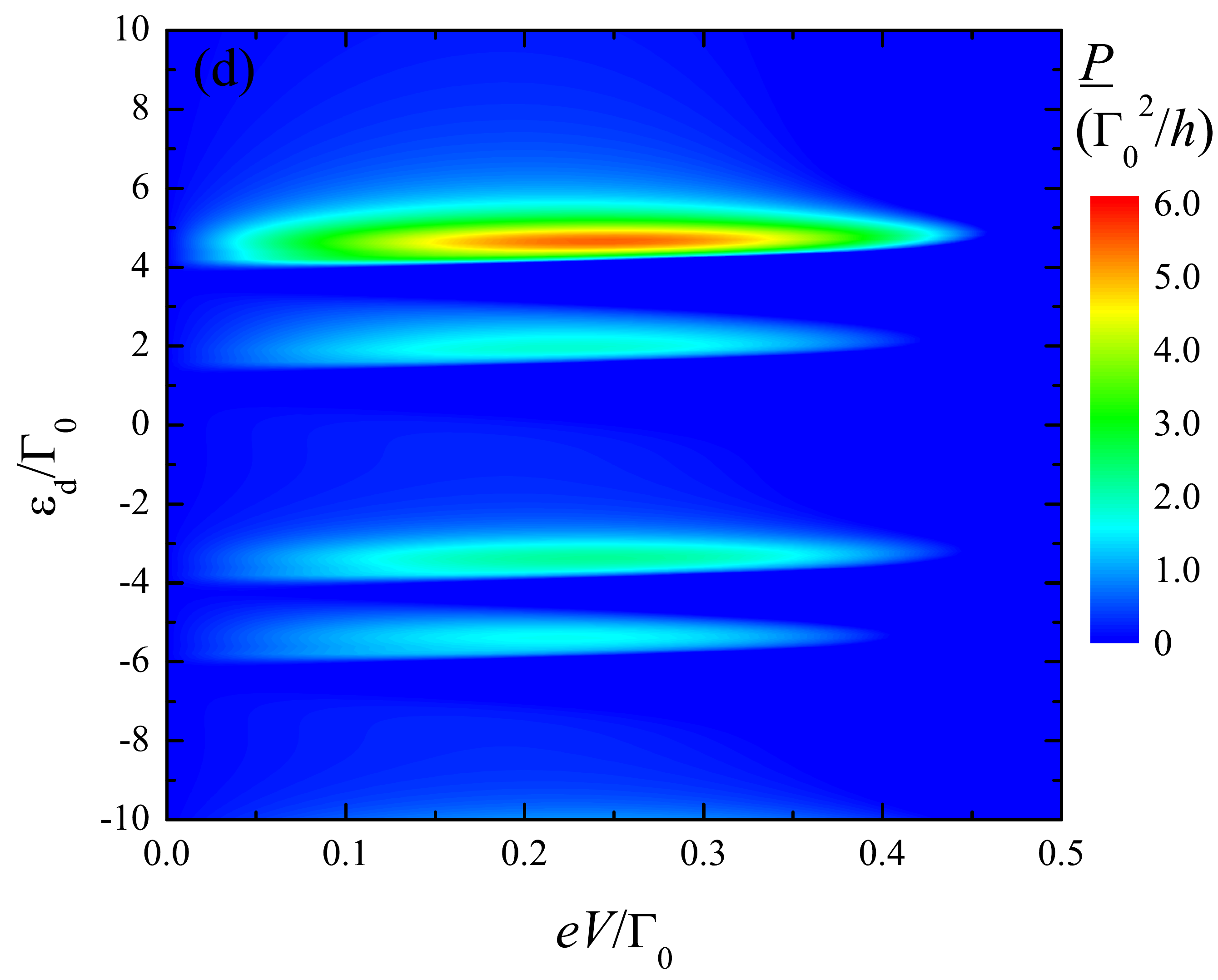}}
\caption{The contour plots display the efficiency (left panels) and the power output (right panels) as a function of the gate potential $\varepsilon_d$ and the bias voltage $eV$ for $k_BT_R=0.0862\Gamma_0$, $k_B\Delta T=0.2\Gamma_0$, $t_c=4\Gamma_0$ and $U=10\Gamma_0$. The upper and lower panels correspond to normal and 1D leads, respectively.}\label{effic0}
\end{figure*}

In order to use this system as a heat engine, we consider the temperature of the left lead $T_L=T+\Delta T$, while the right lead remains at the background temperature $T_R=T$. This generates a voltage $eV=\mu_R-\mu_L$ and a power output $P=IV$. Ideal candidates for efficient heat engines are systems which transmission can be represented by a $\delta$-function. In this case, the system efficiency reaches the Carnot value $\eta=\eta_C$. Unfortunately, the power output goes to zero for this transmission. The problem has been extensively discussed by Hershfield \emph{et al.}\cite{Hershfield} for a non-interacting model. Whitney\cite{Whitney} proposes a boxcar form of the transmission as a candidate that would allow to reach a high efficiency with a large power output.

In the following calculations, we consider $k_BT_R=0.0862\Gamma_0$, $k_B\Delta T=0.2\Gamma_0$ (Carnot efficiency $\eta_C\approx0.7$), $\varepsilon_F=0$, $U=10\Gamma_0$, $\varepsilon_0=\varepsilon_1=\varepsilon_d$ and symmetric couplings to the leads, $\Gamma_{R}=\Gamma_{L}=\Gamma_0/4$.

\begin{figure}
\centerline{\includegraphics[width=88mm,height=73mm,clip]{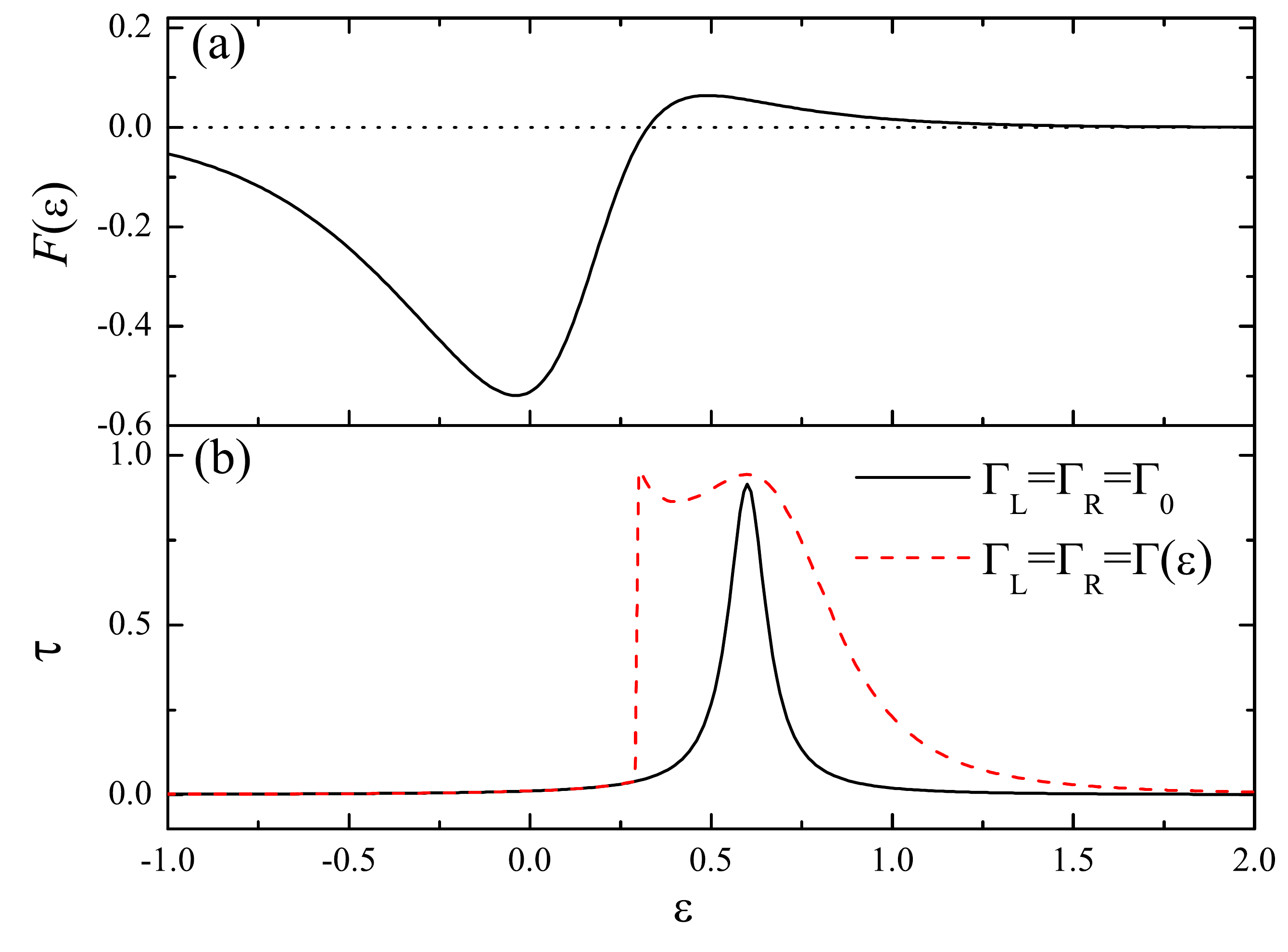}}
\caption{(a) $F(\varepsilon)\equiv f_L(\varepsilon)-f_R(\varepsilon)$ and (b) transmission function for normal (black solid line) and 1D (red dashed line) as a function of the energy for $\varepsilon_d=4.6\Gamma_0$, $t_c=4\Gamma_0$, $k_BT_R=0.0862\Gamma_0$, $k_B\Delta T=0.2\Gamma_0$ and $eV=0.3\Gamma_0$.}\label{effic2}
\end{figure}

Figure \ref{effic0} displays the contour plots for the efficiency (left panels) and power output (right panels) as a function of the gate voltage $\varepsilon_d$ and bias $eV$, for the case of normal (upper panels) and 1D (lower panels) leads. As we can observe in the figure for both kind of leads, the system shows high efficiency and power output in different regions in the parameters space ($\varepsilon_d,eV$). In the case of 1D leads, in these areas, the efficiency and power reach higher values in comparison with the normal leads. In both cases, the efficiency and the power output are optimized when $\varepsilon_d\approx t_c$. As we discussed in the previous section, the enhancement of the efficiency is due to the abrupt change of some relevant quantity around the Fermi energy. In the case of normal leads, the Fano antiresonances produce sudden changes in the transmission, and in the 1D leads case the presence of the Van-Hove singularity in the DOS is responsible for the enhancement of the thermoelectric efficiency. Black regions in panel (a) and (c) correspond to situations where the system receives work from outside and thus the efficiency is greater than the Carnot value. Alternatively, the increase of the power output is characterized by the integrable area of the transmission function in the positive region of the function $F(\varepsilon)\equiv f_L(\varepsilon)-f_R(\varepsilon)$, described in Figure \ref{effic2} (a), which depends on the bias $eV$ and the temperature gradient $\Delta T$. The larger the integrable area in this region, the larger the power output. The appearance of a maximum value in the power output when $\varepsilon_d\approx t_c$, is a consequence of the fact that there is not contribution of the transmission function in the parameter space where $F(\varepsilon)$ takes negative values, i. e., the QDs local levels are all in the positive region of $F(\varepsilon)$. Figure \ref{effic2} (b) shows the transmission function for normal (black solid line) and 1D (red dashed line) leads, around the condition $\varepsilon_d\approx t_c$ where both, efficiency and power output, are enhanced. We see, for normal leads, that for $F(\varepsilon)<0$, the transmission function is essentially zero, existing another peaks in the positive region that correspond to the QDs resonances in $\varepsilon_d+U$. For the case of 1D leads, we observe that the resonance becomes wider increasing the power output, and becomes more abrupt, which enhances the efficiency. It is necessary that the Van-Hove singularity to be located around the energy where the function $F(\varepsilon)$ crosses over from negative to positive values. In this case this energy is $\varepsilon\approx 0.28\Gamma_0$ but in general it can be defined by,
\begin{equation}
\tilde{\varepsilon}=\frac{\mu_RT_L-\mu_LT_R}{T_L-T_R}.
\end{equation}\newline
\begin{figure}
\centerline{\includegraphics[width=88mm,height=123mm,clip]{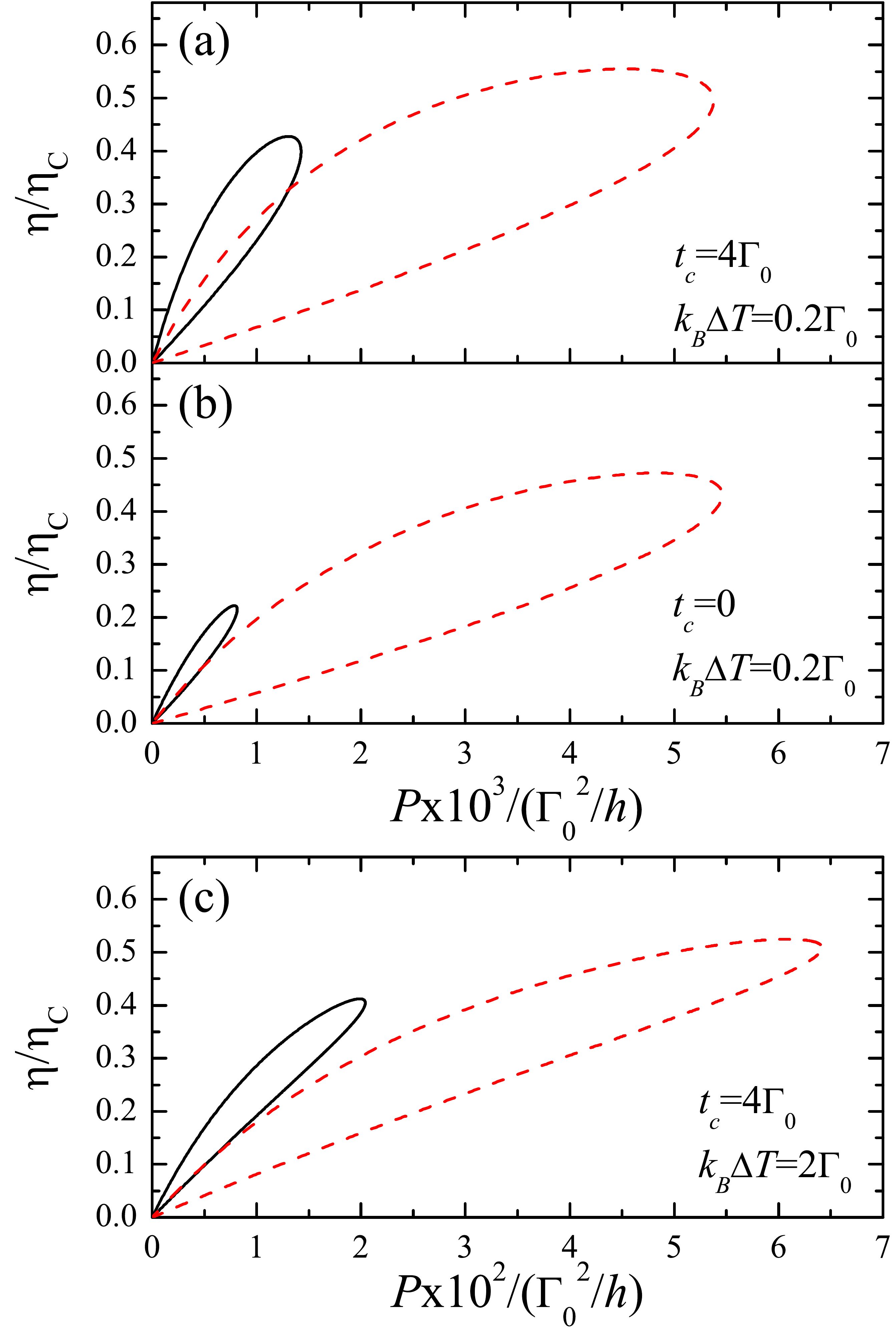}}
\caption{Efficiency as a function of the power output for normal (black solid lines) and 1D leads (red dashed lines) for $k_BT_R=0.0862\Gamma_0$ and $U=10\Gamma_0$. The local level of the dots are $\varepsilon_d=4.6\Gamma_0$ for (a) and (c) panel, and $\varepsilon_d=\Gamma_0$ for (b) panel.}\label{effic1}
\end{figure}

Figure \ref{effic1} (a) shows the efficiency vs power output changing the applied voltage for normal and 1D leads for $\varepsilon_d=4.6\Gamma_0$. This value optimizes both efficiency and power output, almost simultaneously as we can see in Figure \ref{effic0}. For comparison we include, in Figure \ref{effic1} (b), the results for a single QD (i. e. $t_c=0$). For normal leads, it is clear that there is a considerable increase in the efficiency when we connect the side-coupled QD. Nevertheless, for 1D leads, the increase of the efficiency and the power output occurs for both, single and T-shaped QD system. This allow us ti think that a single QD, which is more scalable than the T-shaped configuration, has a great thermoelectric performance when is connected to 1D leads with a Van-Hove singularity near the Fermi level. The Van-Hove singularity transforms the transmission function introducing a very abrupt resonance, but in the case of the T-shaped system, this abrupt change in the transmission is increased by the presence of the Fano resonance originated by the side-coupled QD. Figure \ref{effic1} (c) shows also the efficiency as a function of the power output, modifying the applied potential, for $t_c=4\Gamma_0$ and a larger temperature gradient $\Delta T = 2\Gamma_0$ (Carnot efficiency $\eta_C\approx0.96$). The efficiency shows a little decrease of its maximum value in comparison with panel (a) for normal and 1D leads, however, the power output is increased by an order of magnitude.

\begin{figure}
\centerline{\includegraphics[width=88mm,clip]{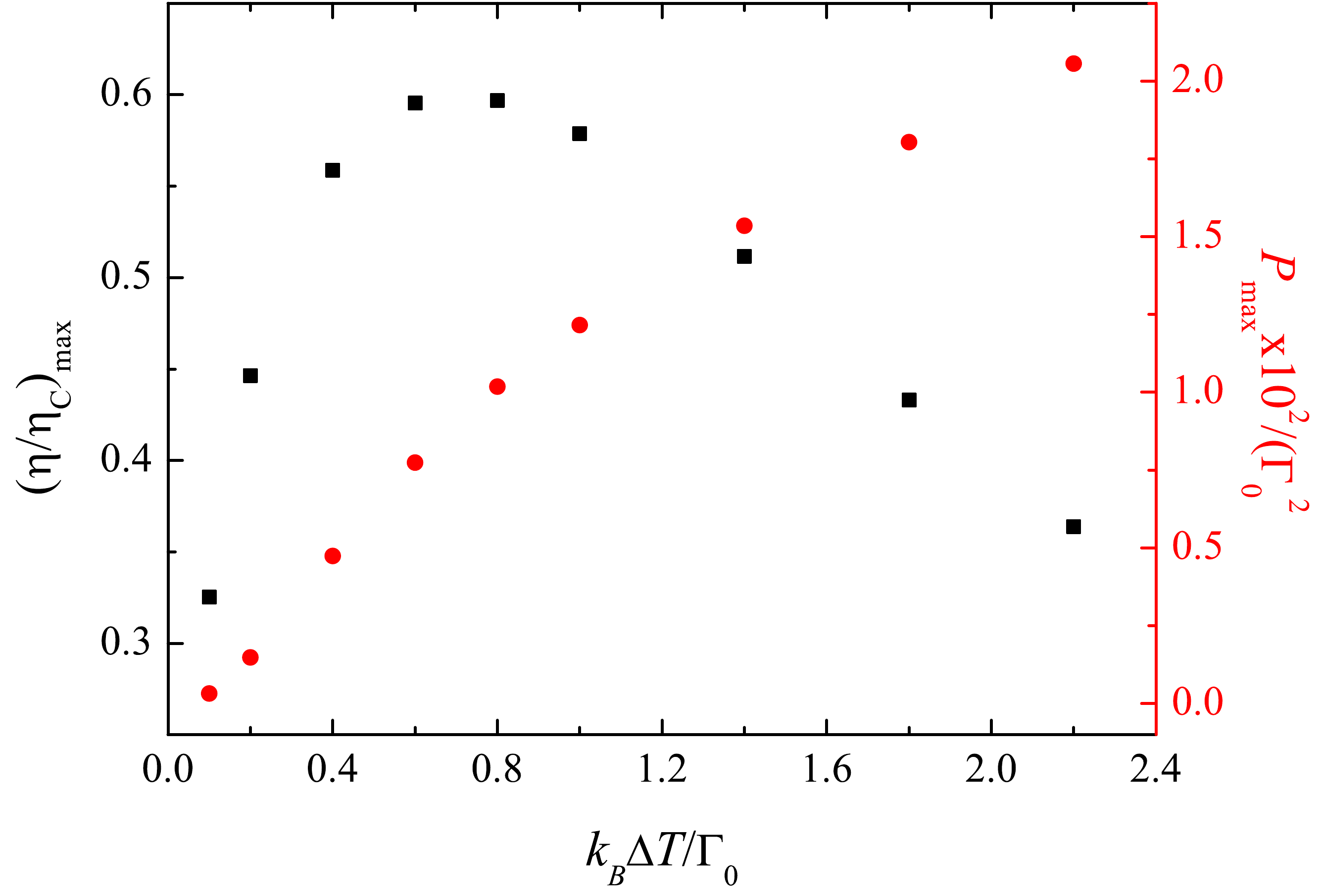}}
\caption{Maximum efficiency (black squares) and maximum power output (red circles) as a function of the temperature gradient for normal leads, $k_BT_R=0.0862\Gamma_0$, $t_c=4\Gamma_0$, $U=10\Gamma_0$ and $\varepsilon_d=4.6\Gamma_0$.}\label{effic3}
\end{figure}

Figure \ref{effic3} shows the maxima values of the efficiency and the power output as a function of the temperature gradient. We observe that the maximum power output grows almost linearly with the temperature gradient while the maximum efficiency has its highest value for $k_B\Delta T\approx0.6\Gamma_0$. Although the positive region of $F(\varepsilon)$ is larger increasing $\Delta T$, which produces an enhancement of the power output, the transmission function becomes smooth and the efficiency drops.

\section*{Summary}

We study the thermoelectric properties of a T-shaped double QD that has been shown to possess high thermoelectric efficiency in the linear and nonlinear regimes. To develop a realistic description of this system, we have incorporated the intra-dot Coulomb repulsion always present in a QD. To do so, we use the Green functions formalism within the Hubbard III approximation, which properly treats the Coulomb blockade regime. The effect of the Coulomb repulsion, because it opens other channels for the electron to go along, reduces the thermoelectric efficiency of the system in both, equilibrium and out of equilibrium conditions. In the nonlinear regime, we carry out a detailed analysis of the thermoelectric efficiency of the system, and we were able to optimize it as a function of the applied voltage and temperature gradient between the leads. We obtain a notable enhancement of the efficiency in comparison with the case with a single QD with normal leads. We have analyzed in detail the case in which the Fermi energy is near a Van-Hove singularity of the contacts DOS. We show that in this case, by adequately manipulating the parameters that define the device, it is possible to obtain a remarkable performance regarding its thermoelectric efficiency. We conclude that the Fano effect and moreover, the Van-Hove singularities near the Fermi energy of contacts with one-dimensional properties are essential ingredients to design a thermoelectric device.

\begin{acknowledgments}
G. G.-S. and E. V. A. acknowledge financial support from the Brazilian agencies CAPES, CNPq and FAPERJ and P. A. O. acknowledges to FONDECYT grant number 1140571.
\end{acknowledgments}

\appendix

\section{Linear conductance in the Hubbard I and Hubbard III approximation}\label{app:hubb}

In this appendix, we study the linear conductance of a system with a strongly correlated region connected to leads using the H$_{\text{I}}$ and H$_{\text{III}}$ approximations. We emphasize the shortcomings derived from the H$_{\text{I}}$ treatment. We consider this an important discussion as the H$_{\text{I}}$ approximation have been extensively used to study the conductance of these type of system for temperatures above the Kondo temperature. For the sake of simplicity, we consider a single QD connected to two metallic leads in the wide band limit. In order to study the conductance,  we calculate the retarded Green function at the QD in the H$_{\text{I}}$ approximation. It is given by,
\begin{equation}
G^I_{00,\sigma}(\varepsilon)=g^I_\sigma(\varepsilon)/(1+\text{i}g^I_\sigma(\varepsilon)\Gamma),
\end{equation}
where,
\begin{equation}
g^I_\sigma(\varepsilon)=\frac{1-\langle n_{\bar\sigma}\rangle}{\varepsilon-\varepsilon_0}+\frac{\langle n_{\bar\sigma}\rangle}{\varepsilon-\varepsilon_0-U},
\end{equation}
and $\Gamma$ represents the self-energy that results from the connection of the QD to the leads and it is supposed to be frequency independent assuming the wide band limit.

The linear conductance at zero temperature, $\mathcal{G}_\sigma(\varepsilon_F)$, is proportional to $-\text{Im}\{G_{00,\bar\sigma}(\varepsilon_F)\}$ and, within this approximation,  can be written as,
\begin{eqnarray}
&&\mathcal{G}^{I}_\sigma(\varepsilon_F)=\\\nonumber
&&\frac{\gamma[\varepsilon_F-\varepsilon_0-U(1-\langle n_{\bar\sigma}\rangle)]^2}{[(\varepsilon_F-\varepsilon_0)(\varepsilon_F-\varepsilon_0-U)]^2+[\varepsilon_F-\varepsilon_0-U(1-\langle n_{\bar\sigma}\rangle)]^2\Gamma^2},
\end{eqnarray}
where, $\gamma=(2e^2/h)\Gamma^2$. In order to study this approximation, we calculate the conductance at resonance, $\varepsilon_0=0$ and $\varepsilon_0=-U$, where we suppose the Fermi level to be $\varepsilon_F=0$. The value for the conductance is the same for these two resonance conditions,
\begin{equation}
\mathcal{G}_\sigma^{I}(\varepsilon_0=0,\varepsilon_F=0)=\mathcal{G}_\sigma^{I}(\varepsilon_0=-U,\varepsilon_F=0)=\frac{\gamma}{\Gamma^2}.
\end{equation}
Surprisingly enough, the conductance does not depend on the occupation number, $\langle n_{\bar\sigma}\rangle$. This is an indication that, within this approximation, the electronic spin $\sigma$ current does not depend upon the QD been charged with electrons with opposite spin. According to this result, all Coulomb blockade effects are eliminated at resonance. The conductance assumes the same value obtained at the one body limit at resonance, $\varepsilon_0=\varepsilon_F$, with no Coulomb repulsion $U=0$. This is obviously an incorrect result. For the sake of comparison we calculate this conductance using the H$_{\text{III}}$ approximation. The Green function $G_{00,\sigma}^{III}(\varepsilon)$ is given by,
\begin{equation}
G_{00,\sigma}^{III}(\varepsilon)=\frac{1-\langle n_{\bar\sigma}\rangle}{\varepsilon-\varepsilon_0+\text{i}\Gamma}+\frac{\langle n_{\bar\sigma}\rangle}{\varepsilon-\varepsilon_0-U+\text{i}\Gamma}.
\end{equation}
Then, we write the conductance as,
\begin{equation}
\mathcal{G}^{III}_\sigma(\varepsilon_F)=\frac{\gamma(1-\langle n_{\bar\sigma}\rangle)}{(\varepsilon_F-\varepsilon_0)^2+\Gamma^2}+\frac{\gamma\langle n_{\bar\sigma}\rangle}{(\varepsilon_F-\varepsilon_0-U)^2+\Gamma^2}.
\end{equation}
Now, we set again the gate voltage at $\varepsilon_0=\varepsilon_F=0$, aligned with the Fermi level, and evaluate the conductance,
\begin{equation}\label{cond:e0}
\mathcal{G}^{III}_\sigma(\varepsilon_0=0,\varepsilon_F=0)=\frac{\gamma(1-\langle n_{\bar\sigma}\rangle)}{\Gamma^2}+\frac{\gamma\langle n_{\bar\sigma}\rangle}{U^2+\Gamma^2}.
\end{equation}
Finally, for the other resonance condition $\varepsilon_0=-U$, the conductance is obtained as,
\begin{equation}\label{cond:eU}
\mathcal{G}^{III}_\sigma(\varepsilon_0=-U,\varepsilon_F=0)=\frac{\gamma\langle n_{\bar\sigma}\rangle}{\Gamma^2}+\frac{\gamma(1-\langle n_{\bar\sigma}\rangle)}{U^2+\Gamma^2}.
\end{equation}
We note that the expressions for the two resonant conductance are not formally equal. However, due to symmetry reasons, the QD occupation $\langle n^1_{\bar\sigma}\rangle$ at $\varepsilon_0=0$ and $\langle n^2_{\bar\sigma}\rangle$ when $\varepsilon_0=-U$ satisfy that $\langle n^1_{\bar\sigma}\rangle+\langle n^2_{\bar\sigma}\rangle=1$, in which case Eqs. \eqref{cond:e0} and \eqref{cond:eU} are equivalent. The result for the conductance, using the H$_{\text{III}}$ approximation, depends on the QD occupation number, reflecting the effect of the Coulomb interaction. This is a fundamental difference compared to H$_{\text{I}}$ approximation.

Finally, we set the gate voltage at the electron-hole symmetry point($\varepsilon_0=-U/2$). The conductance using the H$_{\text{I}}$ is given by,
\begin{equation}
\mathcal{G}^I_\sigma(\varepsilon_0=-U/2,\varepsilon_F=0)=\frac{\gamma(\langle n_{\bar\sigma}\rangle-1/2)^2}{U^2/16+(\langle n_{\bar{\sigma}}\rangle-1/2)^2\Gamma^2}.
\end{equation}
We see that for the electron-hole symmetry condition, ${\langle n_{\bar\sigma}\rangle}=0.5$, the conductance, within this approximation, results to be discontinuous. It assumes the value $\gamma/\Gamma^2$ for $U=0$ and zero for any $U>0$. This is an incorrect result. On the other hand, the H$_{\text{III}}$ conductance is given by,
\begin{equation}
\mathcal{G}^{III}_\sigma(\varepsilon_0=-U/2,\varepsilon_F=0)=\frac{\gamma}{U^2/4+\Gamma^2}.
\end{equation}
We note that the expression is different from zero, and assintotically goes to zero in the limit $U\rightarrow\infty$, which is the qualitatively correct result.

\nocite{*}


\end{document}